\documentclass[aps,reprint,pre,showpacs,superscriptaddress,amsmath,amssymb]{revtex4-1}
\usepackage[utf8]{inputenc}
\usepackage{amsmath}
\usepackage{graphicx}
\usepackage{bm}
\usepackage{color}
\usepackage{algorithm}
\usepackage{algpseudocode}
\usepackage{subfig}

\usepackage{dcolumn}   % needed for some tables
\usepackage{relsize}  % for math
\usepackage{mwe}
\usepackage{float}
\usepackage{mathtools}
\usepackage{amsfonts}
\usepackage{epstopdf}

\begin{document} 
\title{Essentially entropic lattice Boltzmann model: Theory and simulations}
\author{Mohammad Atif}
\affiliation{Jawaharlal Nehru Centre for Advanced 
Scientific Research, Jakkur, Bangalore  560064, India}
\author{Praveen Kumar Kolluru}
\affiliation{Jawaharlal Nehru Centre for Advanced 
Scientific Research, Jakkur, Bangalore  560064, India}
\author{Santosh Ansumali}
\email{ansumali@jncasr.ac.in}
\affiliation{Jawaharlal Nehru Centre for Advanced 
Scientific Research, Jakkur, Bangalore  560064, India} 
\affiliation{SankhyaSutra Labs Limited, Bangalore, India} 
\begin{abstract}
We present a detailed description of the essentially entropic lattice Boltzmann model. 
% [M. Atif, P. K. Kolluru, C. Thantanapally, and S. Ansumali, Phys. Rev. Lett, {\bf 119}, 240602 (2017)].
The entropic lattice Boltzmann model guarantees unconditional numerical stability by iteratively solving the nonlinear entropy evolution equation.
%We reformulate the entropic lattice Boltzmann model to obtain a closed form analytic solution for the discrete path length.
In this paper we explain the construction of closed-form analytic solutions to this equation.
%The essential idea is to relax the entropy equality condition and replace it with the constraint that entropy must increase within a discrete time step. 
We demonstrate that near equilibrium this exact solution reduces to the standard lattice Boltzmann model. 
We consider a few test cases to show that the exact solution does not exhibit any significant deviation from the iterative solution.
We also extend the analytical solution for the ES-BGK model to remove the limitation on the Prandtl number for heat transfer problems.
The simplicity of the exact solution removes the computational overhead and algorithmic complexity associated with the entropic lattice Boltzmann models. 
\end{abstract} 

\maketitle

The  lattice Boltzmann  model (LBM) is an efficient kinetic formulation 
of the nonlinear hydrodynamic phenomena on a lattice designed to capture the physics of macroscopic flow
%in terms  of  a discrete set of  populations restricted on  lattices with appropriate symmetries 
\citep{frisch1986lattice,chen1992recovery,ansumali2003minimal,Wahyu2010,adhikari2005fluctuating,mazloomi2015entropic,kolluru2020extended}.  
The Navier-Stokes dynamics emerges as the hydrodynamic limit of this kinetic model which performs simple microscale
operations on the populations of fictitious particles \citep{higuera1989lattice,qian1992lattice,benzi1992lattice}.
The discrete equilibrium in LBM is chosen such that the macroscopic constraints are satisfied
\citep{mcnamara1988use,qian1992lattice,benzi1992lattice}. 
Historically, the top-down approach of choosing the discrete equilibrium distribution from the macroscopic dynamics 
emerged as a computationally attractive alternative to the Boolean particle dynamics of the lattice gas model \citep{frisch1986lattice,mcnamara1988use, higuera1989lattice}. 
However, this top-down approach lost a few desirable features of the lattice gas such as the unconditional numerical stability, the $H$ theorem
 and consequently the faithful representation of microscopic Boltzmann dynamics  \citep{karlin1999perfect,succi2002colloquium}.   
It was soon realized that the lack of a discrete time $H$ theorem results in the growth of numerical instabilities \citep{boghosian2001entropic,karlin1999perfect,succi2002colloquium}.
%in standard LBM. This often makes simulations  with low viscosity and/or large
% spatial gradients for hydrodynamics and large density ratios for
% multiphase flows unstable \citep{karlin1999perfect,succi2002colloquium,mazloomi2015entropic}.

The entropic lattice Boltzmann model (ELBM) emerged as an alternate methodology to restore the $H$ theorem for discrete space-time evolution
\citep{karlin1998maximum,wagner1998h,karlin1999perfect,chen2000h,boghosian2001entropic,succi2002colloquium,ansumali2003minimal,boghosian2003galilean}.
It was considered a paradigm  shift for computational fluid dynamics because the numerical stability of a hydrodynamic solver was ensured by
compliance with the thermodynamics at the discrete time level \citep{succi2002colloquium}. 
Currently, the ELBM is accepted as a viable tool for simulation of turbulence, multiphase flows, as well as microflows due to its unconditional numerical stability,
and has shown remarkable improvement over the traditional LBM \citep{ANSUMALI2006289,aidun2010lattice,CHIKATAMARLA20131925,mazloomi2015entropic,atif2017}.
The additional step in ELBM, known as the entropic involution step, involves a numerical search for the discrete path length 
corresponding to jump to a mirror state on the isentropic surface.
Considerable efforts have been made to ensure the correctness and efficient implementation of this step 
\citep{ansumali2000stabilization,ansumali2002entropy,tosi2006,chikatamarla2006entropic,brownlee2007stability,gorban2012allowed}. 
However, there is scope for a better theoretical understanding of the ELBM if one is able to obtain a closed form expression for the  discrete path length. For example:
\begin{itemize}
\item The variable discrete path length could be understood as an adaptive implicit modeling of the unresolved scales of the flow via the thermodynamic route, and may provide a new insight into the subgrid modeling of turbulence. 
\item It should enhance the efficiency of the ELBM by avoiding a numerical search for the path length.
\item It will resolve the ambiguities in the implementation of ELBM. It should be noted that for some rare events, the details of which are discussed in Sec. \ref{involutionsec}, the entropic involution step has no solution, and hence there is no unique definition of the path length \citep{gorban2012allowed}.
\end{itemize}

In Ref. \citep{atif2017}, the authors reformulated the ELBM and obtained a closed form analytical solution for the discrete path length $\alpha$.
This was achieved by relaxing the entropy equality condition used in ELBM and replace it with the constraint that entropy must increase within a discrete time step. 
The analytical form of $\alpha$ was found as the root of a quadratic equation $-a\alpha^2 + b \alpha -c $, 
% and written as
% \begin{equation}
% \alpha = \frac{-b + \sqrt{b^2-4ac } }{-2a}, 
% \end{equation}
where the coefficients $a,b,c$ are given in Eq. \eqref{qaud_tight_coeff}.
The near equilibrium limit of this exact solution is the standard LBGK value of $\alpha=2$. 
Its simplicity removes the computational overhead and algorithmic complexity associated with ELBM. 
In this paper, we discuss the theory of the entropic lattice Boltzmann model and explain the construction of the  closed form analytic solution for the discrete path length in detail.
We also demonstrate that the exact solution exhibits no significant deviation from the iterative ELBM solution by considering a few canonical setups.
%The proposed methodology is devoid of indeterminacy due to its exact nature.
This paper is organized as follows: In Sec. \ref{ELBMintro}, we briefly review the entropic lattice Boltzmann model. 
In Sec. \ref{involutionsec}, we describe the entropic involution step in its traditional form and derive its near-equilibrium limit. 
In Sec. \ref{eelbmsec}, we explain the methodology to construct exact solutions for the path length. 
In Sec. \ref{comparisonOfElbm}, we perform a detailed comparison of the our solution with ELBM and BGK values of path length.
% In Sec. \ref{secnaca0012} we present the simulation of a NACA0012 airfoil at Reynolds number $2.88\times 10^6$, and 
In Sec. \ref{esbgk} we derive the analytical solution to the path length for the ES-BGK model. Finally, in Sec. \ref{subvisc} we derive the expression for turbulent viscosity corresponding to the exact solution of the path length.

\section{Entropic lattice Boltzmann model} \label{ELBMintro}

In this section, we introduce the LBM and its entropic formulation in $D$ dimensions. 
In LBM one defines a set of  discrete velocities  ${\bf c}_i$, $i=1,\cdots, N$ such that they form links of a space-filling lattice \citep{succi2001lattice}, and at every lattice node ${\bm x}$ and time $t$ a set of discrete  populations $f({\bm c}_i, {\bm x},t)\equiv f_i$. Here, the set of populations $f_i$ is understood as a vector $\bm f=\{f_1,f_2,\cdots,f_N\}$ in the $N$ dimensional vector space, where $N$ is the number of discrete populations.
We define the bilinear action between two functions  of discrete velocities $\phi$ and $\psi$ as 
\begin{equation}
\left< \phi, \psi\right>=\sum_{i=1}^N \phi_i \psi_i.
\end{equation}
Analogous to continuous kinetic theory, the hydrodynamic variables such as the mass density $\rho$, velocity $\mathbf{u}$, and the scaled temperature $\theta$ are defined
as
\begin{equation}
 \rho=\left< f, 1\right>, \quad  \rho {\bm u}= \left< f,  {\bm c}\right>, \quad 
 \rho u^2 +   D \rho \theta = \left< f,  {\bm c}^2\right>.
\end{equation} 
Similarly, the $H$ function for hydrodynamics is taken in Boltzmann form as
\citep{karlin1999perfect,ansumali2003minimal,ansumali2005consistent}  
\begin{equation}
H[f] = \left<f, \log\frac{f}{w}-1\right>,
\label{hdef}
\end{equation} 
with weights $w_i>0$. 
The population $\bm f( {\bm x} + {\bm c}_i \Delta t, t+\Delta t)$ after a time step $\Delta t$ 
starting from $\bm f({\bm x},t)$ is written as two step process:
\begin{enumerate}
\item The discrete free-flight as
\begin{equation}
\bm f( {\bm x} + {\bm c}_i \Delta t, t+\Delta t) = \bm f^{*}({\bm x},t),
\label{advection}
\end{equation}
which shifts the populations from one lattice node to another. Similar to the free flight of molecules, this step preserves the entropy globally, i.e., $\sum_{\bm x} H[f( {\bm x} + {\bm c}_i \Delta t, t+\Delta t)] = \sum_{\bm x} H[f]$ (see Ref. \citep{wagner1998h} for a detailed proof).

\item The collisional relaxation towards the discrete equilibrium as
\begin{equation}
\label{fstar_def}
 \bm f^{*}({\bm x},t) = \bm f({\bm x},t) + \alpha \beta \left[\bm f^{\rm eq}(\mathcal{M}^{\rm slow}({\bm x},t))-\bm f({\bm x},t) \right],
\end{equation} 
typically modeled by a single relaxation model of Bhatnagar-Gross-Krook (BGK) \citep{bhatnagar1954model} with  mean free time $\tau$.
Here, $\mathcal{M}^{\rm slow}({\bm x},t) = \{\rho({\bm x},t), \bm u({\bm x},t), \theta({\bm x},t) \}$ are the collisional invariants ($\theta({\bm x},t) \notin \mathcal{M}^{\rm slow}({\bm x},t)$ for isothermal LBM).
For the standard LBGK, $\alpha=2$, and the dimensionless discrete relaxation parameter $\beta = {\Delta t }/{ (2\tau+\Delta t) }$ is bounded in the interval $0<\beta<1$. Notice that $\beta=1$ implies $\tau=0$, and as the kinematic viscosity $\nu  = \tau \theta$, $\beta=1$ implies that there is no dissipation in the system. 
For a typical LBM simulation the operating range is an over-relaxation regime of $\Delta t/\tau \gg 1$ where $\beta \rightarrow 1$. In the standard LBM, this regime of $\beta \rightarrow 1$ encounters numerical instability, which is resolved in the ELBM by treating $\alpha$ as a variable which is evaluated at each point and time step such that the $H$ theorem is satisfied. This is discussed in detail in Sections \ref{involutionsec}-\ref{eelbmsec}.  
\end{enumerate}

To recapitulate, the discrete free-flight that represents the convection process leads to no dissipation, hence no entropy production \citep{wagner1998h}.
%some remark about entropy flux
The collisional relaxation, however, has non-zero entropy production due to relaxation of the populations towards the equilibrium but is entirely local in position space. 

Historically, the discrete isothermal equilibrium at a reference temperature $\theta_0$ was chosen as \citep{qian1992lattice}
\begin{equation}
f_i^{\rm eq} = w_i \rho \left[1 + \frac{u_\alpha c_\alpha}{\theta_0} +  \frac{u_\alpha u_\beta}{2\theta_0^2} \left(c_\alpha c_\beta - \theta_0 \delta_{\alpha\beta} \right)\right],
\label{histeq}
\end{equation}
which was sufficient to recover the Navier-Stokes dynamics upto $\mathcal{O}(u^2)$, provided that the moments of the weights $w_i$ satisfy
\begin{equation}
\left<w, 1 \right> = 1, \, \left<w, c_\alpha c_\beta \right> = \theta_0 \delta_{\alpha\beta}, \left<w, c_\alpha c_\beta c_\gamma c_\kappa \right> = \theta_0^2 \Delta_{\alpha\beta\gamma\kappa}, 
\end{equation}
where $\Delta_{\alpha\beta\gamma\kappa} =  \delta_{\alpha\beta} \delta_{\gamma\kappa} + \delta_{\alpha\gamma} \delta_{\beta\kappa} + \delta_{\alpha\kappa} \delta_{\beta\gamma} $.
However, this polynomial form of discrete equilibrium permits the populations to attain negative values thus making the simulations numerically unstable \citep{karlin1999perfect,succi2002colloquium}. 
A method that resolves the issue of nonpositive form of equilibrium distribution is to construct the discrete equilibrium $\bm f^{\rm eq}$ as the minimizer of the convex $H$ function under the constraint that the mass density, the momentum density, and the energy density (ignored for isothermal scenarios) are conserved \citep{karlin1999perfect,boghosian2001entropic,atif2018,kolluru2020lattice}.
The discrete entropic equilibrium thus obtained is of the form
\begin{equation}
 f_i^{\rm eq} = w_i \rho \exp\left(-\mu - \zeta_\alpha c_{i\alpha} -\gamma c_i^2 \right),
 \label{entropiceqdef}
\end{equation}
where $\mu,\zeta_\alpha,\gamma$ are the Lagrange multipliers. For the $D1Q3$ model, the discrete entropic isothermal equilibrium in the explicit form is
\begin{equation}
f_{\pm 1}^{\rm eq} = \frac{\rho }{6} \,\varUpsilon \left[\frac{2 {u}_\alpha + \sqrt{1+3{u}_\alpha^2}}{1-{u}_\alpha} \right]^{\pm 1}, 
\quad f_0^{\rm eq} = \frac{4\rho}{6} \, \varUpsilon,
% , \quad f_{1}^{\rm eq} = \frac{\rho}{6}  \, \varUpsilon \left[\frac{2\tilde{u}_x/\sqrt{3} + \sqrt{1+\tilde{u}_x^2}}{1-\tilde{u}_x/\sqrt{3}} \right],
\label{d1q3eq}
\end{equation}
where $\varUpsilon = 2-\sqrt{1+3{u}^2}$.
For the higher-dimensional extensions of $D1Q3$, i.e., $D2Q9,D3Q27,$ the generalized expression of the discrete entropic isothermal equilibrium is
\citep{ansumali2003minimal}
\begin{equation}
f_i^{\rm eq} = w_i \rho \prod_{\alpha=1}^D \varUpsilon \left[\frac{ 2{u}_\alpha + \sqrt{1+3{u}_\alpha^2}}{1-{u}_\alpha } \right]^{c_{i\alpha}/\sqrt{3\theta_0}}.
\end{equation}
The above entropic equilibrium can be compared with Eq. \eqref{histeq} by performing a series expansion around $u=0$. The expansion up to ${\cal O}(u^3)$ is   
\begin{align}
\begin{split}
f_i^{\rm eq} = w_i \rho \bigg[ 1 + \frac{u_\alpha c_\alpha}{\theta_0} +  \frac{u_\alpha u_\beta}{2\theta_0^2} \left(c_\alpha c_\beta - \theta_0 \delta_{\alpha\beta} \right) \\+ \frac{1}{6\theta_0^3} \left( u_\alpha u_\beta u_\gamma c_\alpha c_\beta c_\gamma - \theta_0 u^2 u_\alpha c_\alpha \right) \bigg],
\end{split}
\end{align}
which matches the historically employed equilibrium from Eq. \eqref{histeq} till ${\cal O}(u^2)$. The errors in the higher moments such as viscous stress and heat flux is of ${\cal O}(u^4)$ and ${\cal O}(u^3)$ respectively \citep{ansumali2004minimal}.
As for most higher-order models, the Lagrange multipliers cannot be evaluated in explicit form and need to be found numerically. The series form can be used as an alternative for simulations at low Mach numbers (${\rm Ma}$) defined as ${\rm Ma} = u/c_s $, where $c_s$ is the sound speed.  

\section{The entropic involution} \label{involutionsec}

The existence of the entropy function $H$ accompanied with the entropic equilibrium derived in a variational fashion provides an opportunity for creating a nonlinearly stable numerical method \citep{karlin1999perfect,succi2002colloquium,boghosian2001entropic}.
As the advection process [Eq. \eqref{advection}] does not lead to entropy production \citep{chen2000h}, a nonlinearly stable LBM can be achieved by making the collisional relaxation to equilibrium [Eq. \eqref{fstar_def}] adhere to the $H$ theorem, or in other words, by ensuring that there is nonpositive entropy production during the collision \citep{karlin1999perfect}.

The physical domain is discretized into grid points, at each of which we define a set of $N$ populations $\bm f = \{f_0, f_1, \cdots f_{N-1} \}$. 
Each point has an entropy level $H$ associated with it. For example, at a grid point with set of populations $\bm f^+ = \{f^+_0, f^+_1, \cdots f^+_{N-1} \}$, from Eq. \eqref{hdef}, $H[f^+]$ is a scalar. 
The equilibrium $\bm f^{\rm eq}$ is the point with the least value of $H$, as, by construction, it is the minimizer of the convex entropy function $H$ under the relevant constraints.

\begin{figure}
\centering
    \includegraphics[width=0.5\textwidth]{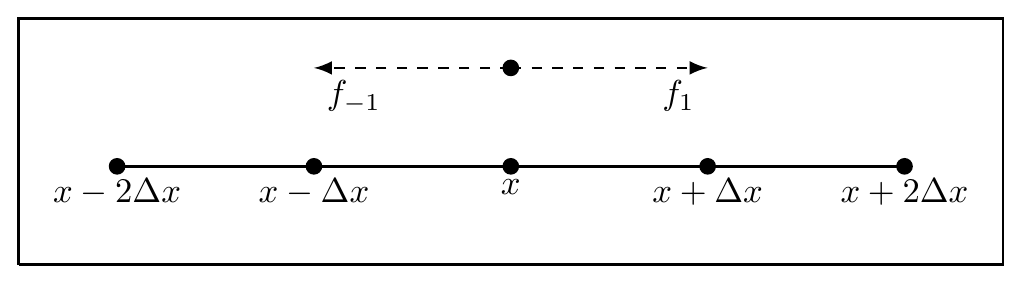}
    \caption{Discrete velocities in a D1Q2 model. This one dimensional lattice has only two populations $f_1, f_{-1}$ with discrete velocities $+1, -1$ respectively, and cannot model hydrodynamics due  to the lack of enough degrees of freedom.}
    \label{twovelmodel}
\end{figure}

\begin{figure}
\centering
    \includegraphics[width=0.43\textwidth]{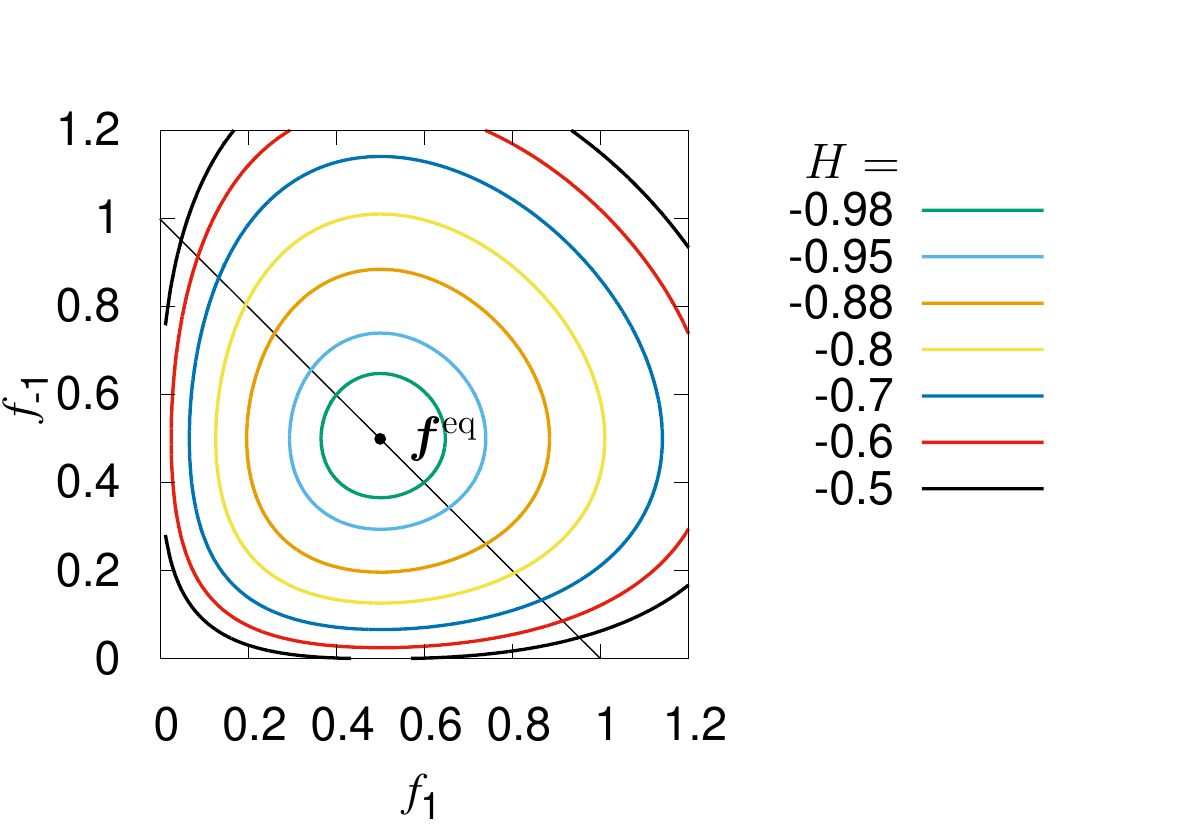}
    \includegraphics[width=0.43\textwidth]{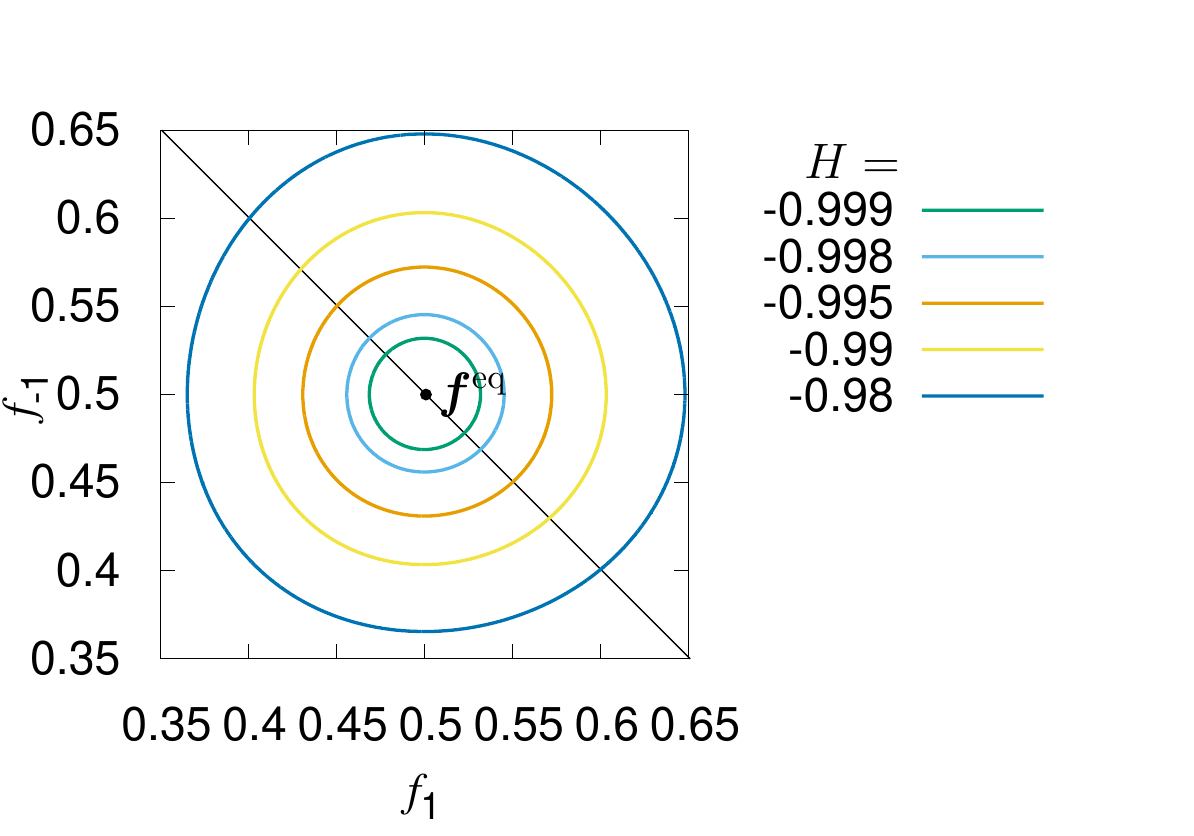}
    \caption{Isoentropy contours for a $D1Q2$ lattice. It can be seen from zoomed figure (bottom) that near the equilibrium the isoentropy contours become almost circular.}
    \label{convexentropyd1q2}
\end{figure}

The collision step given by Eq. \eqref{fstar_def} is understood in geometric terms as follows: in an $N$ dimensional phase space, starting from the pre-collisional state $\bm f$, one covers a distance (path length) $\alpha\beta$ in the direction of $\bm f^{\rm eq} - \bm f$ to reach the post-collisional state $\bm f^*$, i.e.,
\begin{equation}
\bm f^* = \bm f  +\alpha\beta [ \bm f^{\rm eq} - \bm f ].
\label{fstardef2}
\end{equation}
Here, for convenience we have dropped the position and time coordinates $\bm x,t$ as the collision step is local in position space and instantaneous.
We first consider the $D1Q2$ lattice as an example to visualize the phase space and discuss the entropic collisional dynamics. 
This one dimensional lattice has only two populations $f_1, f_{-1}$ with discrete velocities $+1, -1$ respectively (see Fig. \ref{twovelmodel}). Due to the lack of enough degrees of freedom, the $D1Q2$ lattice does not conserve momentum and hence cannot model hydrodynamics. 
The mass density ($\rho = f_1 + f_{-1}$) is a conserved moment, and the momentum density ($\rho u = f_1 - f_{-1}$) becomes a nonconserved moment.
These two constraints can be inverted to obtain the relations
\begin{equation}
f_1 = \frac{\rho + \rho u}{2} , \quad f_{-1} =  \frac{\rho - \rho u}{2}. 
\label{d1q2mominvert}
\end{equation}
Figure \ref{convexentropyd1q2} represents the isoentropic contours in the vector space for the $D1Q2$ lattice.
The criterion of mass conservation $f_1 + f_{-1} = \rho$ dictates that the collisional dynamics for $\rho=1$ is restricted on the straight line in the figure. The equilibrium is given by
\begin{equation}
{\bm f}^{\rm eq} \equiv \{f_1^{\rm eq}, f_{-1}^{\rm eq} \} = \{\rho/2,\rho/2\}. 
\end{equation}
It can be seen from the Fig. \ref{convexentropyd1q2} (bottom) that near the equilibrium the isoentropy contours are almost circular. This property of the $H$ function $ (H = f_1 \log f_1 + f_2 \log f_2 - \rho - \rho \log 2)$  is valid for the higher dimensional lattices as well.

\begin{figure}
\centering
{\includegraphics[width=0.35\textwidth]{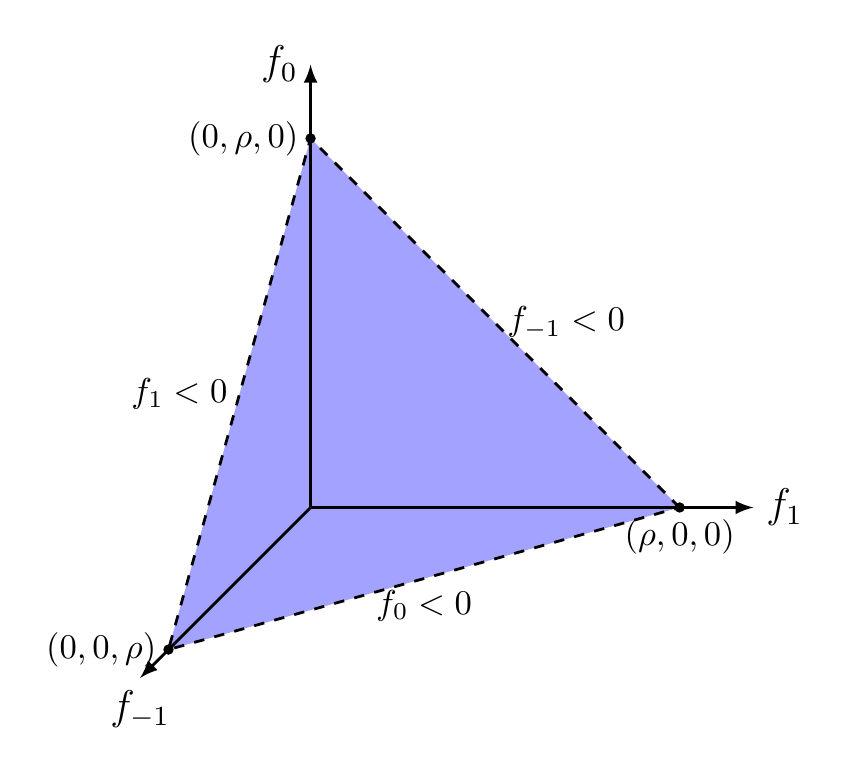}}
{\includegraphics[width=0.35\textwidth]{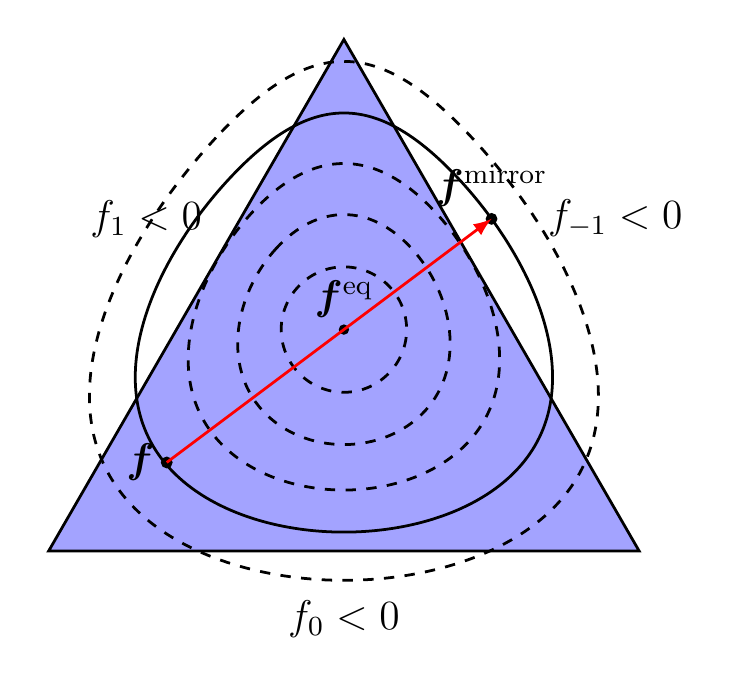}}
\caption{Top: The polytope of positivity for the $D1Q3$ lattice is a triangular section of the plane inside which all the populations are positive, and outside of which one or
more populations become negative. Bottom: Representation of a pre-collisional state $\bm f$ for which the mirror state is not defined.}
\label{elbm_issue}
\end{figure}

Another model we consider is D1Q3, which will be used later for illustrating the concepts of entropic involution.
For the $D1Q3$ lattice, the populations are $\{f_{-1},f_0,f_1\}$ with discrete velocities $\{-1,0,+1\}$ respectively.
The mass conservation constraint requires that $f_{-1}+f_0+f_1=\rho$, a plane on which the entire discrete dynamics is constrained (see Fig. \ref{elbm_issue}).
The equilibrium  for the $D1Q3$ lattice is given by Eq. \eqref{d1q3eq}.
The conserved moments are the mass density  $\rho = f_{-1}+f_0+f_1$ and momentum density $\rho u = f_1-f_{-1}$, whereas the nonconserved moment is the stress $\sigma_{xx} = f_1+f_{-1}-f^{\rm eq}_1-f^{\rm eq}_{-1}$. These three constraints can be inverted to obtain the relations
\begin{align}
\begin{split}
&\tilde f_{-1}\equiv \frac{f_{-1}}{\rho} = \frac{ \tilde f^{\rm eq}_1+\tilde f^{\rm eq}_{-1} + \tilde\sigma_{xx} - u}{2}, \\
&\tilde f_0   \equiv \frac{f_0   }{\rho} = 1 - \tilde\sigma_{xx} -\tilde f^{\rm eq}_1-\tilde f^{\rm eq}_{-1} , \\
&\tilde f_1   \equiv \frac{f_1   }{\rho} = \frac{ \tilde f^{\rm eq}_1+\tilde f^{\rm eq}_{-1} + \tilde\sigma_{xx} + u}{2},
\label{d1q3faseq}
\end{split}
\end{align}
where $\tilde\sigma_{xx} = \sigma_{xx}/\rho, \tilde f^{\rm eq}_i = f^{\rm eq}_i/\rho$.

\begin{figure}
\centering
    \includegraphics[width=0.3\textwidth]{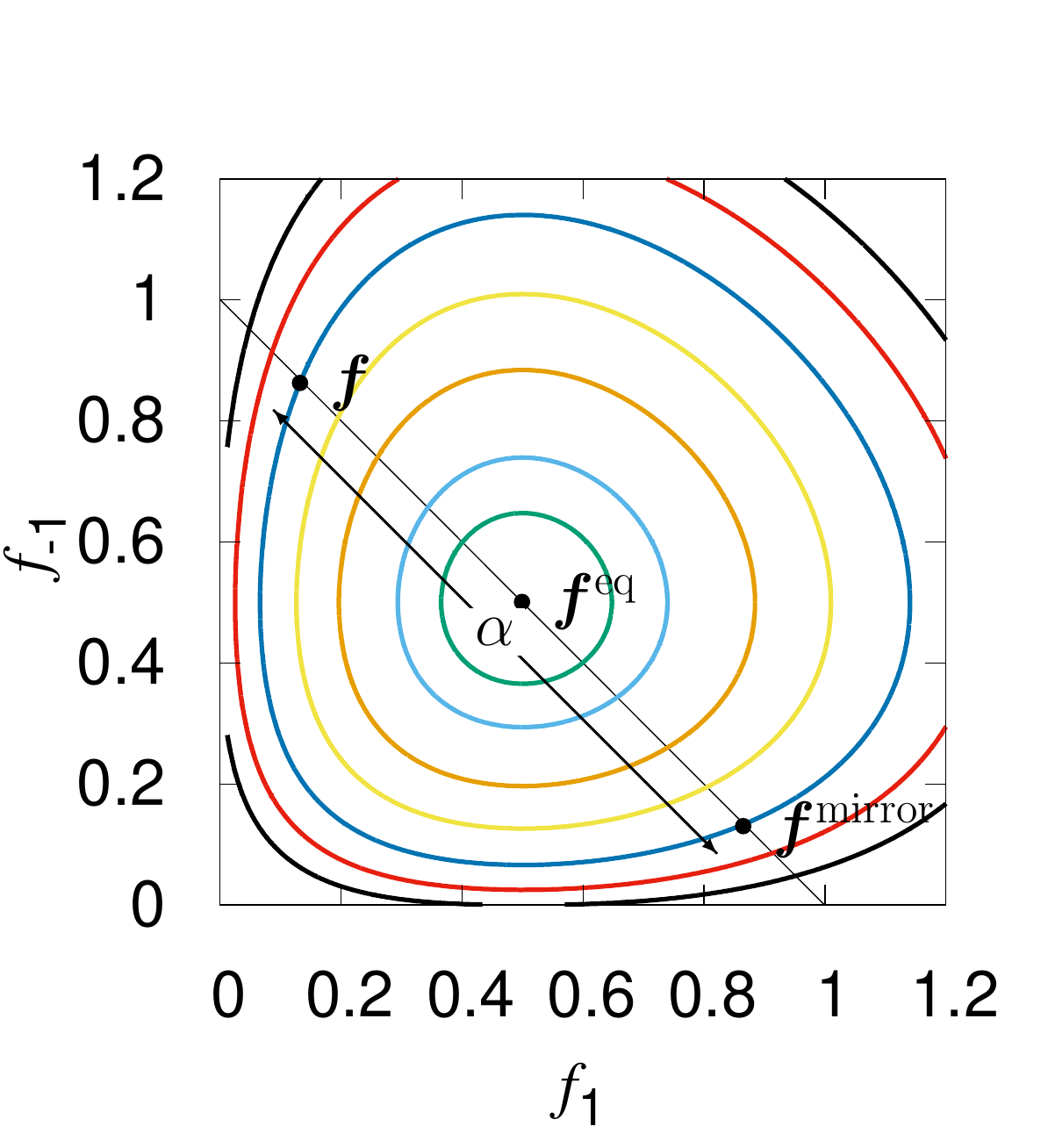}
    \caption{Entropic collisional dynamics for $D1Q2$ lattice. Note that the pre-collisional state $\bm f$ and the mirror state ${\bm f}^{\rm mirror}$ are at the same entropy level.}
    \label{mirror_d1q2}
\end{figure}

%In the LBGK, $\alpha=2$, however, in ELBM $\alpha$ deviates from its standard value of 2 and is found such that the $H$ theorem is satisfied. 
We now define a mirror state 
\begin{equation}
\bm f^{\rm mirror} = \bm f +\alpha ( \bm f^{\rm eq} - \bm f ) ,
\label{fmirrordef}
\end{equation}
which is essentially $\bm f^*$ from Eq. \eqref{fstardef2} with $\beta=1$. 
Here, we remind that $\beta=1$ is a zero dissipation state, therefore, the mirror state $\bm f^{\rm mirror}$ lies at the same entropy as the initial state $\bm f$, i.e.,
\begin{equation}
H[\bm f^{\rm mirror}] = H[\bm f].
\label{entropy_conserve}
\end{equation}
The aim of the entropic involution step is to find the $\alpha$ corresponding to the mirror state.
Note that all the states $\bm f, \bm f^*, \bm f^{\rm mirror}$ are at a higher entropy level than $\bm f^{\rm eq}$.
Hence, starting from $\bm f$ and moving in the direction of $\bm f^{\rm eq} - \bm f$, the value of $H$ decreases till the equilibrium state, after which it begins to rise.
The maximum allowable path length that could be covered is $\alpha$, after which $H$ increases beyond its pre-collisional state, and the $H$ theorem is violated. This is depicted in Fig. \ref{mirror_d1q2} for the D1Q2 lattice.

There exists an important structure in the distribution functions space -- the polytope of positivity \citep{gorban2012allowed}.
It is the region inside which all the populations are positive but outside of which one or more populations become negative. 
The shaded triangular region in Fig. \ref{elbm_issue} (top) is the polytope of positivity for the $D1Q3$ lattice. 
The entropic involution does not yield a solution when the isoentropic surfaces are partially outside the polytope of positivity.
This is due to the presence of the logarithm in the entropy function which is undefined when one of the populations is negative. 
Figure \ref{elbm_issue} (bottom) shows a pre-collisional state $\bm f$ for which the mirror state lies outside the triangle, hence cannot be defined.

In LBGK, the path length is fixed to a constant value of $\alpha_{\rm LBGK}=2$.
The ELBM introduces the concept of the state dependent $\alpha$ \citep{karlin1999perfect}, evaluated numerically by solving the nonlinear equation [Eq. \eqref{entropy_conserve}] \citep{ansumali2002entropy,tosi2006,chikatamarla2006entropic}.
Once the path length $\alpha$ and therefore the mirror state are known, the post-collisional state is found by the linear contraction
\begin{equation}
\bm f^* = \bm f^{\rm mirror} - \alpha(1-\beta)[\bm f^{\rm eq} - \bm f] = \bm f + \alpha\beta[\bm f^{\rm eq} - \bm f].
\end{equation}
Since $0 < \beta < 1$, it is guaranteed that $H[\bm f^*] < H[\bm f^{\rm mirror}]$.
To summarize, the ELBM ensures adherence to the $H$ theorem in the collision by first ``over-relaxing'' the populations to an equal entropy (zero dissipation) mirror state followed by adding dissipation, thus, ensuring a nonpositive entropy production \citep{karlin1999perfect}. 

Next, we discuss the near equilibrium limit of the entropic involution.
In a well resolved simulation, the departure of populations from the equilibrium is small and the entropic involution step yields the solution $\alpha=\alpha_{\rm LBGK}=2$. 
To demonstrate this, we define the dimensionless departure from the equilibrium as
\begin{align}
\label{xdef}
x_i=\frac{f_i^{\rm eq}}{f_i}-1.
\end{align}
As the populations $f_i, f_i^{\rm eq}$ are positive, $x_i \in (-1,\infty)$. 
Here, the lower limit is due to the extreme case of $f_i^{\rm eq} \rightarrow 0$, whereas the upper limit is due to $f_i \rightarrow 0$.
Further, we introduce a decomposition of distributions $f_i$ in terms of the departure from equilibrium as \citep{gorban1996relaxational}
\begin{equation}
\Omega^{+}=\{f_i:x_i \geq 0\}, \quad \Omega^{-}=\{f_i: -1<x_i<0\}. 
\label{decompose}
\end{equation}
This asymmetry in the range of $x$ is crucial in the subsequent derivation of the exact solution. 
With this decomposition, we also partition the bilinear action into two partial 
contributions  
\begin{equation}
\left<f,\psi \right>_{\Omega^\pm} = \sum_{f_i \in \Omega^\pm} f_i \psi_i.
\end{equation}

The path length $\alpha$ is the root of the equation
 \begin{align}
\Delta H \equiv H[\bm f^{\rm mirror}]-H[\bm f] =0,
\end{align}
which is simplified to obtain (see Appendix \ref{derivationdeltaH} for a detailed derivation)
\begin{align}
\begin{split}
H[\bm f^{\rm mirror}]-H[\bm f] = \left<  f, \left( 1+  \alpha  x \right) \log{  \left( 1+  \alpha  x\right)} \right> \\-\alpha  \left< f, x \log(1+x) \right>.
\label{deltaH_}
\end{split}
\end{align}
In a well resolved simulation, the dimensionless departure of populations from the equilibrium is small, i.e.,  $|x_i| \ll 1$.
Therefore, expanding the above equation about $x_i=0$ via Taylor series  one obtains 
\begin{equation}
H[\bm f^{\rm mirror}]-H[\bm f] = \alpha \left(\frac{\alpha}{2}-1 \right) \left<f,x^2 \right> + O(x^3).
\end{equation}
Thus, for small departure from the equilibrium, the non-trivial root of $H[\bm f^{\rm mirror}]-H[\bm f]=0$ is $\alpha = 2$. 
Hence, in the limit $x_i \rightarrow 0$, the ELBM reduces to the LBGK.\\
 
\begin{figure}
\centering
    \includegraphics[width=0.45\textwidth]{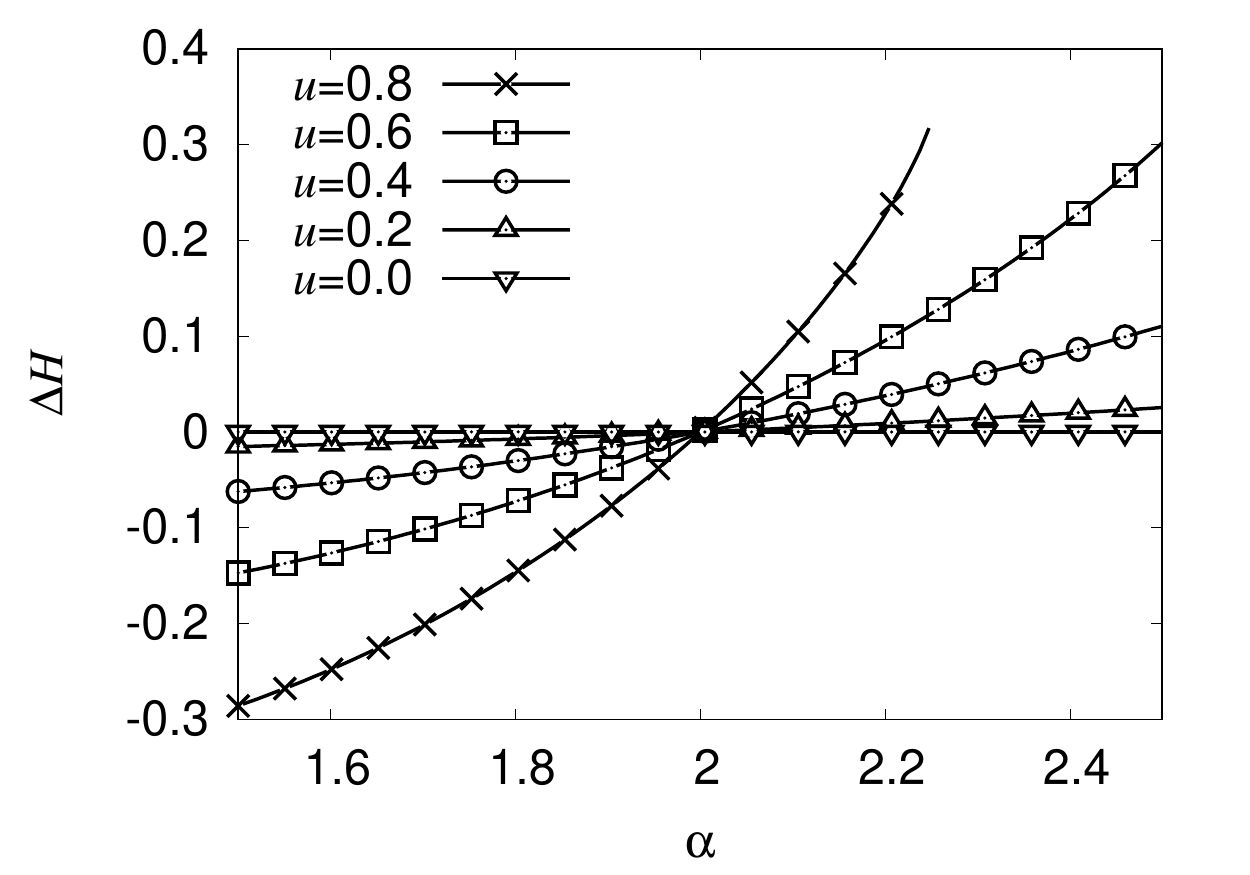}
    \caption{The solution for $\Delta H = 0$ remains $\alpha=2$ at all values of $u$ for the $D1Q2$ lattice (this is not the case for $D1Q3$ and other higher lattices).}
    \label{d1q2alpha2}
\end{figure}

We now derive the expanded form of Eq. \eqref{deltaH_} for the $D1Q2$ lattice. As stated earlier, the $D1Q2$ lattice lacks the degrees of freedom to model hydrodynamics, however, it is simple enough to show the analytical form of $H[\bm f^{\rm mirror}]-H[\bm f]$. 
The Eq. \eqref{deltaH_} for the $D1Q2$ lattice can be expanded to obtain
\begin{align}
\begin{split}
&\Delta H \equiv H[\bm f^{\rm mirror}]-H[\bm f] \\&=  f_{1} (1+\alpha x_{1}) \log(1+\alpha x_{1}) - \alpha f_{1} x_{1} \log(1+ x_{1}) 
\\& + f_{-1} (1+\alpha x_{-1}) \log(1+\alpha x_{-1})  - \alpha f_{-1} x_{-1} \log(1+ x_{-1}).
\end{split}
\end{align}
For this lattice, $f^{\rm eq}_{1} = f^{\rm eq}_{-1} = \rho/2$, therefore, $x_{1} = \rho/(2f_{1})-1, x_{-1} = \rho/(2f_{-1})-1$, substituting which in the above  equation along with Eq. \eqref{d1q2mominvert} yields
\begin{align}
\begin{split}
&\frac{\Delta H}{\rho}  = \left[ \frac{1 + u - \alpha u}{2}  \right] \log \left[ \frac{1+ u -\alpha u}{ 1+ u } \right] 
 \\&+ \left[ \frac{1 - u + \alpha u}{2} \right] \log \left[  \frac{ 1 - u + \alpha  u}{ 1 - u } \right] 
+ \frac{\alpha u}{2} \log \left[ \frac{1-u}{1+u} \right].
\end{split}
\end{align}
It is seen from the above equation that the solution of $\Delta H = 0$ is independent of $\rho$. It can also be verified that $\alpha=2$ is a nontrivial solution (this is due to the symmetric nature of $D1Q2$ and is not the case for $D1Q3$ and other higher dimensional lattices). 
Figure \ref{d1q2alpha2} shows that the solution for $\Delta H = 0$ remains $\alpha=2$ at all values of $u$. 
% However, the complexity associated with the above equation makes it difficult to solve it numerically. 

\begin{figure}
\centering
    \includegraphics[width=0.5\textwidth]{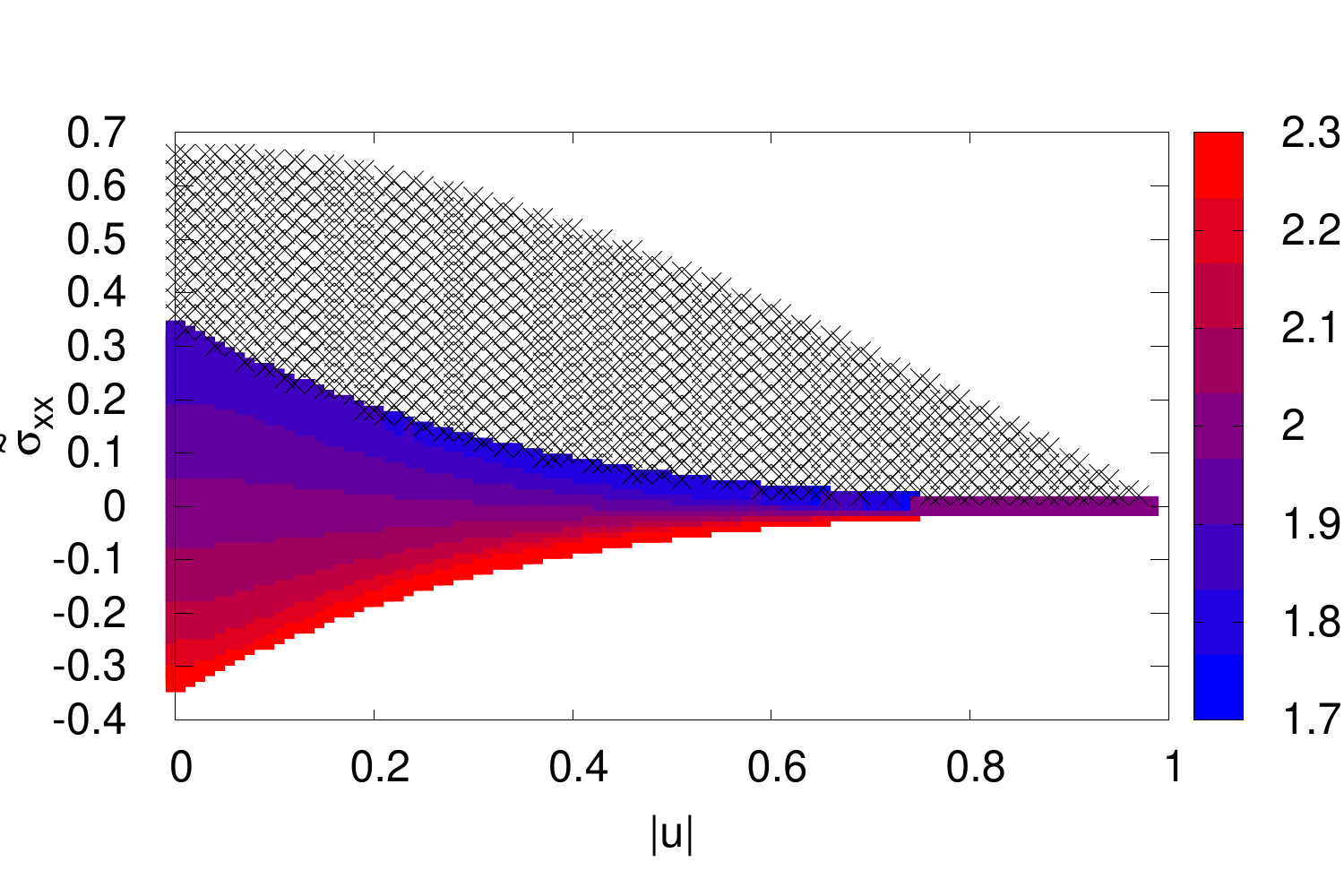}
    \includegraphics[width=0.45\textwidth]{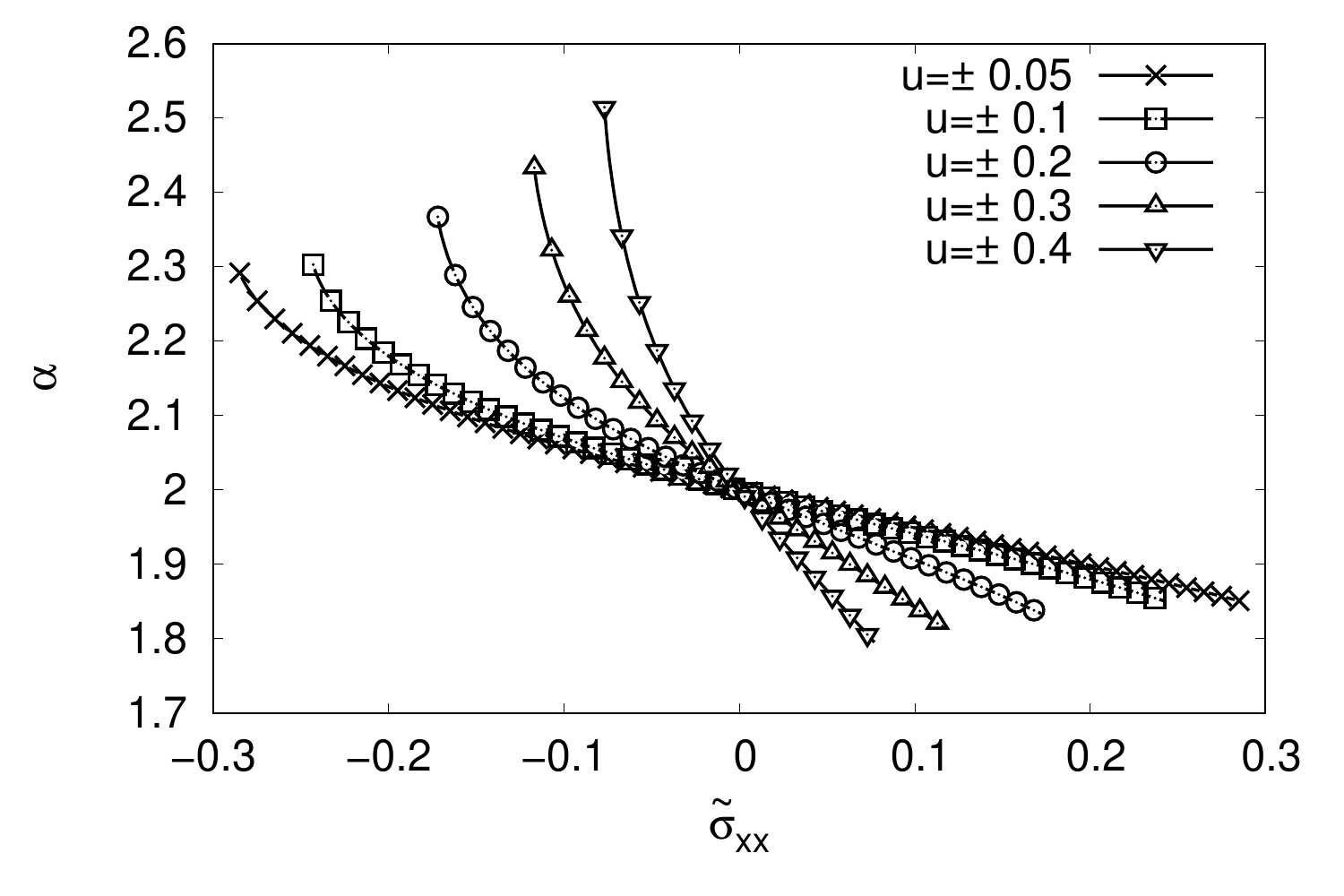}
    \caption{The heat map of $\alpha$ corresponding to $\Delta H = 0$ at various values of $u, \tilde\sigma_{xx}$ for the $D1Q3$ lattice. The shaded region represents the part of moment space where the mirror state lies outside the polytope of positivity.}
    \label{d1q3alpha}
\end{figure}

Next, we derive the expanded form of Eq. \eqref{deltaH_} for the $D1Q3$ lattice.
We define $\tilde x_i$ as the $x_i$ for the $D1Q3$ model which are calculated by substituting the equilibrium from Eq. \eqref{d1q3faseq} into Eq. \eqref{xdef} as
\begin{align}
\begin{split}
&\tilde x_{-1} = \frac{2 \tilde f^{\rm eq}_{-1} }{ \tilde f^{\rm eq}_1+\tilde f^{\rm eq}_{-1} + \tilde\sigma_{xx} - u} -1, \\
&\tilde x_0     = \frac{\tilde f^{\rm eq}_0}{1 - \tilde\sigma_{xx} -\tilde f^{\rm eq}_1-\tilde f^{\rm eq}_{-1}} -1, \\
&\tilde x_1     = \frac{2 \tilde f^{\rm eq}_{1} }{ \tilde f^{\rm eq}_1+\tilde f^{\rm eq}_{-1} + \tilde\sigma_{xx} + u}-1.
\end{split}
\end{align}
The above $\tilde x_i$ are substituted in Eq. \eqref{deltaH_} to obtain the entropy evolution for $D1Q3$ as 
 \begin{align}
 \begin{split}
 &\frac{\Delta H}{\rho} =  \tilde f_{1} \left[ (1+\alpha \tilde x_{1}) \log(1+\alpha \tilde x_{1}) - \alpha \tilde x_{1} \log(1+\tilde x_{1}) \right] 
 \\
  & +\tilde f_{-1} \left[ (1+\alpha \tilde x_{-1}) \log(1+\alpha \tilde x_{-1}) - \alpha \tilde x_{-1} \log(1+\tilde x_{-1}) \right] 
  \\
 & +\tilde f_{0} \left[ (1+\alpha \tilde x_{0}) \log(1+\alpha \tilde x_{0}) - \alpha \tilde x_{0} \log(1+\tilde x_{0}) \right], 
 \end{split}
 \end{align}
which is then solved using Newton-Raphson scheme for the path length $\alpha$. This path length is dependent on $\tilde \sigma_{xx}$ and $u$ of the initial state $\bm f$. Figure \ref{d1q3alpha} plots the values of $\alpha$ for various $u,\tilde \sigma_{xx}$. It can be seen that the region corresponding to the LBGK value of 2, becomes thinner as $|\bm u|$ increases, and that the deviation of $\alpha$ from the LBGK value becomes larger as $|\bm \tilde{\sigma}_{xx}|$ increases. Figure \ref{d1q3alpha} (bottom) plots the path length as a function of $\bm \tilde{\sigma}_{xx}$ for various values of the velocity $|\bm u|$.
The shaded portion of the Fig. \ref{d1q3alpha} (top) represents the regions (typically with large moments) where the initial state is well defined (lies within the polytope of positivity), whereas the mirror state lies outside the polytope of positivity, thus, for such cases, the entropic involution shows indeterminacy.
It should be noted that these events are rare and even if one encounters such cases it is known how to construct the path length \citep{ansumali2002entropy,mazloomi2015fsi}.

We now discuss the significance of over-relaxation in the entropic involution step over the under-relaxation. 
A numerical scheme via the first order Euler discretization of the Boltzmann BGK equation is possible. It reads as
\begin{align}
f(\bm x + \bm c \Delta t, t+\Delta t) &= f(\bm x,t) + \frac{\Delta t}{\tau}\left[f^{\rm eq} -  f(\bm x, t) \right] \notag\\
&=\left(1- \frac{\Delta t}{\tau}\right) f(\bm x,t) + \frac{\Delta t}{\tau}f^{\rm eq},
\end{align}
and exhibits unconditional numerical stability if $\Delta t \ll \tau$. 
The $H$ theorem for this scheme is trivially satisfied as the post-collisional state is a convex combination of the pre-collisional state and the equilibrium state.
This is called an under-relaxing scheme as the discrete dynamics never crosses over the equilibrium state and corresponds to $\alpha<1$.
%However, at large Reynolds number, the viscosity is small and hence $\tau \rightarrow 0$.
%As the under-relaxation requires $\Delta t \ll \tau$, it also sets a severe restriction on the time step, 
%i.e., $\tau \rightarrow 0 \Rightarrow \Delta t \rightarrow 0$. 
However, for many practical applications the relevant time scales are multiple orders of magnitude greater than $\Delta t$.
Therefore, for faster convergence it is required to have numerical scheme which permits large time steps, i.e., $\Delta t \gg \tau$ is desirable (which correspond to $\alpha>1$).
% In Figs. \ref{mirror_d1q2} and \ref{elbm_issue}, all the points on the line segment ${\bm f}^{\rm mirror} - {\bm f}^{\rm eq}$ are called over-relaxed states, while those on the line segment ${\bm f}^{\rm eq} - {\bm f}$ are called under-relaxed. 
The over-relaxation of the populations to a mirror state is thus an important feature of the discrete dynamics as it allows one to achieve large time steps. 

\section{Exact solution to the path length: Essentially entropic lattice Boltzmann model} \label{eelbmsec}

As discussed in the previous section, the discrete path length $\alpha$ is available as the nontrivial root of Eq. \eqref{deltaH_}.
This equation is highly nonlinear and is typically solved by a combination of bisection and Newton-Raphson method \citep{ansumali2000stabilization,ansumali2002single}. 
Considerable efforts have been put in to ensure that the correct solution is obtained in an efficient manner \citep{ansumali2002entropy,tosi2006,chikatamarla2006entropic,brownlee2007stability}.
In this section, we present an alternate construction of ELBM where the discrete path length $\alpha$ is known in explicit form without any indeterminacy. 
The key idea is to obtain $\alpha$ by directly considering the natural criterion of monotonic decrease of $H$ with time \citep{atif2017}. This
implies solving an inequality 
\begin{equation}
\Delta H \equiv H[\bm f^{*}] -H[\bm f] < 0.
\label{entropy_decrease}
\end{equation}
The above inequality, by  construction, accepts multiple solutions. 
For example, when $\alpha\leq1$ the inequality is trivially satisfied as the new state is a convex combination of the old state and the equilibrium \citep{wagner1998h}. 
However, one is interested in an over-relaxed collision, where the new state 
 is no longer a convex combination of the old state and equilibrium. This corresponds to  
 the real solutions of Eq. \eqref{entropy_decrease} in the range $1 < \alpha < \alpha^{\rm max}$, where  $\alpha^{\rm max} = -1/\left(\beta x_i^{\rm min} \right)$ is maximum possible pass-length corresponding to an edge of the polytope of positivity beyond which the populations become negative \citep{karlin1999perfect}. 
Among the multiple solutions of the inequality, we are looking for the maximal path length $\alpha$ such that ${\Delta H \rightarrow 0 }$.
As is the case with ELBM, the solution should reduce to
standard LBM close to equilibrium ($\alpha=2$). Indeed, the present methodology is valid for both discrete velocity models of LBM as well as the continuous in velocity Boltzmann-BGK equations, where the  summation in the inner products needs to be replaced by appropriate integrals.

The general idea behind obtaining an analytical expression for the path length $\alpha$ is as follows:  we intend to split $\Delta H$ into two parts,
\begin{equation}
\Delta H = H(\alpha) + H^{(B)},
%\Delta H = g(\alpha) + \sigma^{(B)},
\label{genidea}
\end{equation}
where $H^{(B)}$ is chosen such that it is nonpositive, and $H(\alpha)=0$ is an easily solvable polynomial whose root is the path length $\alpha$. 
The discrete-time $H$ theorem is satisfied as $H^{(B)}$ is nonpositive and contributes to the entropy production, i.e.,
\begin{equation}
\Delta H =  H^{(B)} \leq 0.
\end{equation}
A word of caution is in order here. As stated earlier, the inequality $\Delta H \leq 0$ by construction accepts multiple solutions.
These solutions are not identical but differ in two ways:
\begin{enumerate}
\item Not all the solutions reduce to LBGK $(\alpha_{\rm LBGK}=2)$ in the limit of $x_i \rightarrow 0$. Our interest is only in the solutions that reduce to the standard LBM for $x_i \rightarrow 0$.
\item The entropy production corresponding to each solution dictates its dissipative nature, i.e., as the magnitude of $H^{(B)}$ increases the dynamics becomes more and more dissipative. 
This is the reason why we are interested in the solution such that ${\Delta H \rightarrow 0 }.$
This point will be elucidated in the forthcoming section, where we derive two expressions for $\alpha$, one of which is more dissipative than the other. 
\end{enumerate} 

Following the procedure detailed in Appendix \ref{derivationdeltaH} the Eq. \eqref{entropy_decrease} is rewritten as
 \begin{align}
%  \begin{split}
 \Delta H =& \left< f, \left( 1+  \hat x \right)\log{  \left( 1+  \hat x\right)} \right> -\alpha \beta \left< f, x \log(1+x) \right>,
\label{deltaH3}
% \end{split}
\end{align}
where $\hat x = \alpha\beta x$. Under the decomposition given by Eq. \eqref{decompose}, the above equation becomes
 \begin{align}
 \begin{split}
 \Delta H &= \left< f, \left( 1+  \hat x \right)\log{  \left( 1+  \hat x\right)} \right>_{\Omega^{-}} + \left< f, \left( 1+  \hat x \right)\log{  \left( 1+  \hat x\right)} \right>_{\Omega^{+}}  \\&-\alpha \beta \left< f, x \log(1+x) \right>_{\Omega^{-}} -\alpha \beta \left< f, x \log(1+x) \right>_{\Omega^{+}}.
\label{deltaH4}
\end{split}
\end{align}
We now derive two solutions to $\Delta H \leq 0$ by splitting Eq. \eqref{deltaH4} into a polynomial and an entropy production term as in Eq. \eqref{genidea}.
These solutions require bounds on the logarithm. The lower order solution is constructed by exploiting the loose bounds, whereas the higher order solution is derived by exploiting the sharper bounds (see Appendix \ref{logbounds} for details on the bounds of logarithm). Both the solutions are shown to reduce to the LBGK value of 2 for $x_i \rightarrow 0$. 

\subsection{Lower order solution} \label{lowerordersoln}

In this section, we find the path length by exploiting the loose bounds on the logarithms [Eqs. \eqref{G1bound},\eqref{G2bound},\eqref{G3bound}].
Upon adding and subtracting the term $\left<f, {\cal A}_1 + {\cal A}_2 - {\cal A}_3 \right>$ from Eq. \eqref{deltaH4}, it is written as 
\begin{align}
\begin{split}
\Delta H = \left<   f,  \left( 1+  \hat x \right) \log{  \left( 1+  \hat x \right)}- {\cal A}_1 \right>_{\Omega^{-}} 
+ \left<   f, {\cal A}_1   \right>_{\Omega^{-}} \\+ \left<  f, {  \left( 1+  \hat x \right)
     \log{  \left( 1+  \hat x \right)}- {\cal A}_2 } \right>_{\Omega^{+}}
     + \left<  f,  {\cal A}_2 \right>_{\Omega^{+}}  \\-\alpha \beta  \left< f, { x \log(1+x) - {\cal A}_3 }  \right> - \alpha \beta  \left< f, {\cal A}_3 \right>,
\label{deltaHexpanded}
\end{split}
\end{align}
where
\begin{align}
\begin{split}
{\cal A}_1 = \hat x + \frac{ \hat x^2}{2}   - \frac{ \hat x^3}{2},\,
{\cal A}_2 = \hat x  +\frac{ \hat x^2}{2},\,
{\cal A}_3 = \frac{2x^2}{2+x}.
\end{split}
\end{align}
Now, identifying that $\left<f,x \right>_{\Omega^{+}}+\left<f,x \right>_{\Omega^{-}}= \left<f,x \right> = 0$ due to conservation laws, Eq. \eqref{deltaHexpanded} 
is written in a compact form as
\begin{align}
\Delta H &= \alpha \beta H_1(\alpha) + H_1^{(B)},
\label{deltaHcompact}
\end{align}
where
\begin{align}
\begin{split}
H_1^{(B)} = - {\left<f,G_1(\hat x) \right>_{\Omega_-}}
 - {\left<f,G_2(\hat x) \right>_{\Omega_+}} 
 -{\alpha\beta\left<f,G_3(x) \right>} \\
  + {\alpha^2 \beta (\beta-1) \left<f,\frac{x^2}{2}\right>} 
- { \alpha^3\beta(\beta^2-1)\left< f, \frac{x^3}{2} \right>_{\Omega^{-}} },
\label{sigmaBlower}
\end{split}
\end{align}
and
\begin{equation}
H_1(\alpha)=  -\alpha^2 a_1 +\alpha b_1  - c_1 ,
\label{dissipativescheme}
\end{equation}
with
\begin{equation}
\quad a_1=\left<f, \frac{x^3}{2} \right>_{\Omega^{-}}, b_1=\left< f, \frac{x^2}{2} \right>, c_1 = \left< f, \frac{2x^2}{2+x} \right> .
\label{alphaLowerCoeffs}
\end{equation}
It can be seen that $H_1(0)<0<H_1(2)$, therefore, a positive root of Eq. \eqref{dissipativescheme} bounded in $(0,2)$ exists.
As Eq. \eqref{dissipativescheme} is constructed by employing lower order bounds on the logarithm, this root is called $\alpha_{\rm Lower}$,
\begin{equation}
\alpha_{\rm Lower}=\frac{ -b_1 +  \sqrt{ b_1^2 - 4a_1 c_1 } }{ -2a_1 } 
= \frac{ 2c_1 }{ b_1 +  \sqrt{ b_1^2 - 4a_1 c_1} }.
\label{alphaLowerSoln}
\end{equation}
To avoid numerical issues related to the precision loss while dealing with small numbers, in the above expression we have multiplied the root with its conjugate \citep{press1992numerical}.

Due to the nonnegative nature of the functions $G_1, \, G_2, \, G_3$ in their respective domains [Eqs. \eqref{G1bound}, \, \eqref{G2bound}, \,\eqref{G3bound}], and 
$\beta<1$, each term in Eq. \eqref{sigmaBlower} is nonpositive, hence, 
$H_1^{(B)} \leq 0$.
Therefore, from Eq. \eqref{deltaHcompact} we see that the $H$ theorem is satisfied because $H_1(\alpha_{\rm Lower}) = 0$, hence,
\begin{align}
\Delta H =  H_1^{(B)}  \leq 0.
\label{compactLower}
\end{align}
Upon expanding $\alpha_{\rm Lower}$ and ignoring higher order terms one obtains 
\begin{equation}
\lim_{x_i \rightarrow 0} \alpha_{\rm Lower} = 2 - \frac{ \left< f, x^3 \right>_{\Omega^{+}}}{ \left< f, x^2 \right>} + 3  \frac{ \left< f, x^3 \right>_{\Omega^{-}}}{ \left< f, x^2 \right>}, 
\label{limitalpha}
\end{equation}
which has the limiting value of $2$. 
Thus, for small departures from equilibrium where $x_i \rightarrow 0$, the scheme reduces to the standard LBM. 
It is also evident from Eq. \eqref{limitalpha} that $\alpha_{\rm Lower}<2$.
This is important as it is known that for ELBM the path length fluctuates around the standard LBGK value of $\alpha=2$ \citep{karlin2015entropy},
a feature of ELBM not mimicked by $\alpha_{\rm Lower}$. In the next section, we construct another path length $\alpha_{\rm Higher}$
that fluctuates about the standard LBGK value of $\alpha=2$.

\subsection{Higher order solution}
In this section, we derive the path length $\alpha$ by exploiting the sharper bounds on the logarithms [Eqs. \eqref{G4bound}, \, \eqref{G5bound}, \, \eqref{G6bound}]. 
Following the same methodology as the previous section, we add and subtract terms from Eq. \eqref{deltaH} to obtain
\begin{align}
\begin{split}
\Delta H &= H^{(B)}+ \alpha\beta H(\alpha), 
\end{split}
\end{align} 
where $H^{(B)}<0$ and 
\begin{align}
H(\alpha) = -\alpha^2 a + \alpha b - c.
\label{qaud_tight} 
\end{align}
The coefficients $a,b,c$ are
\begin{widetext}
\begin{align}
\begin{split}
a &=  \beta^2 \left<   f, \frac{ x^3}{6} - \frac{ h \beta x^4}{12} + \frac{ h^2 \beta^2 x^5}{20} - \frac{ h^3 \beta^3 x^6}{5} \right>_{\Omega^{-}}, c = \left< f, \frac{60x^2+60x^3+11x^4}{60+90x+36x^2+3x^3} \right>, \\
b &=  \bigg<f, \frac{ x^2}{2}  \bigg> - \bigg<f, \frac{ 2 \alpha_{\rm Lower} \beta^2 x^3}{15}\bigg(
\frac{2}{ 4+\alpha_{\rm Lower} x } + \frac{1}{ 4+2 \alpha_{\rm Lower} x } 
+ \frac{2}{ 4 +3 \alpha_{\rm Lower} x}   \bigg) \bigg>_{\Omega^{+}},\\
\end{split}
\label{qaud_tight_coeff}
\end{align}
\end{widetext}
The parameter $h$ in the above equation serves as an upper bound on the path length and is found as the positive root of the quadratic equation  
\begin{align}
\begin{split}
H_2(\alpha) &= -\alpha^2 a_2 + \alpha b - c,
\end{split}
\label{gaplhah0}
\end{align}
where
\begin{align}
\begin{split}
a_2 = \beta^2 \left<   f, \frac{ x^3}{6} \right>_{\Omega^{-}}.
\end{split}
\end{align}
Equation \eqref{qaud_tight} has a positive root  $\alpha_{\rm Higher}$ [as $H(0)<0<H(\infty)$] which is the desired path length. 
It has the limit
\begin{equation}
\lim_{x_i \rightarrow 0} \alpha_{\rm Higher} = 2 + \left( \frac{4\beta^2}{3}-1\right) \frac{\left<f,x^3\right>}{\left<f,x^2\right>}. 
\label{alphahigherlimit}
\end{equation}
Unlike $\alpha_{\rm Lower}$, which was always less than 2, no such comment can be made about $\alpha_{\rm Higher}$. Thus, $\alpha_{\rm Higher}$ mimics an important feature of the
ELBM where the path length fluctuates about the BGK value of 2.
A detailed derivation of $\alpha_{\rm Higher}$ is provided in Appendix \ref{higherordersoln}.
The details regarding the implementation of this exact solution for the path length are given in Appendix \ref{implementing}.

\section{Comparison with ELBM and BGK} \label{comparisonOfElbm}

In this section, we compare the analytical solutions for the path length ($\alpha_{\rm Lower}$,  $\alpha_{\rm Higher}$) with the BGK ($\alpha_{\rm LBGK}=2$) and the iterative ELBM solution ($\alpha_{\rm ELBM}$).
To this end, we consider three canonical setups: the one-dimensional Sod shock tube, the doubly periodic shear layer, and the lid-driven cavity.
It is illustrated from these examples that $\alpha_{\rm Lower}$ is more dissipative than $\alpha_{\rm Higher}$ and hence is not the ideal choice for hydrodynamics.
Nevertheless, it is useful for the construction of $\alpha_{\rm Higher}$ as demonstrated in the previous section.
It is also demonstrated that there is an insignificant difference between the path lengths  $\alpha_{\rm Higher}$ and $\alpha_{\rm ELBM}$.

\subsection{Sod shock tube}

\begin{figure*}
{\includegraphics[width=0.99\textwidth]{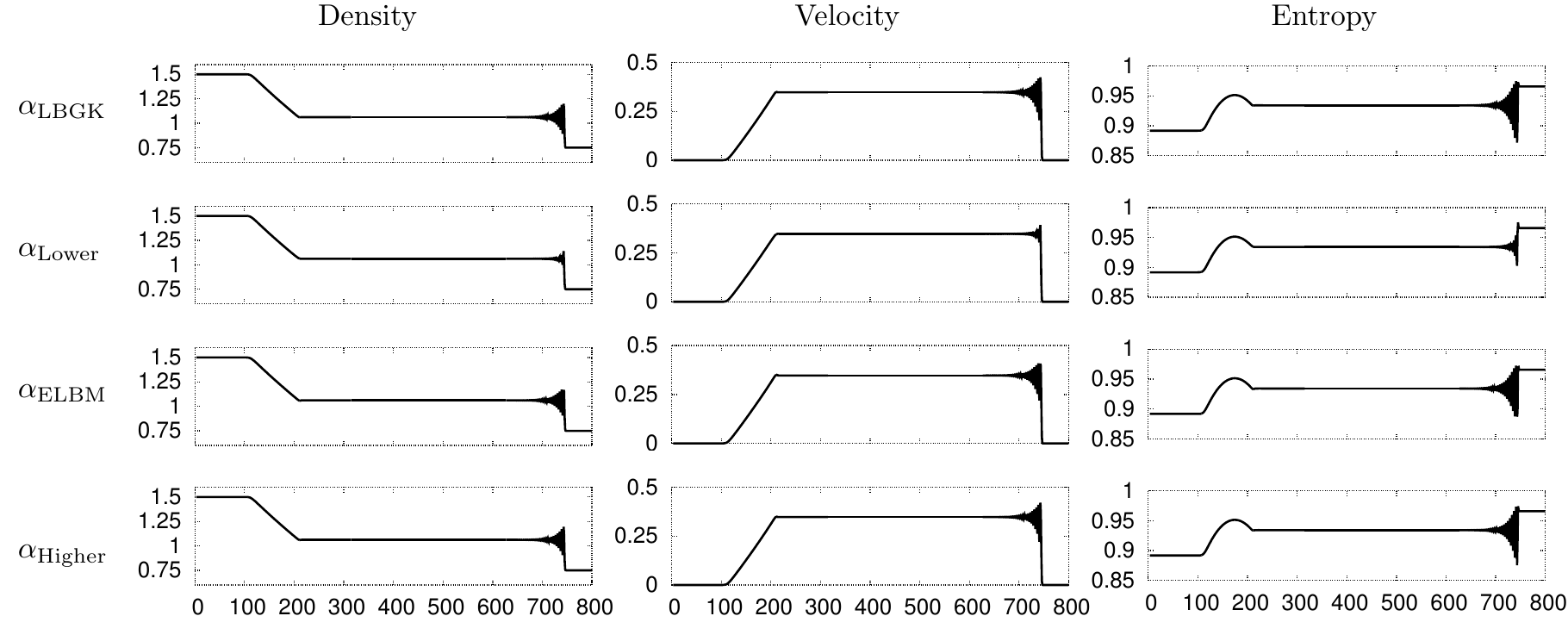}}
    \caption{Density (left), velocity (middle) and entropy (right) plots from $\alpha_{\rm LBGK}$, $\alpha_{\rm Lower}$, $\alpha_{\rm Higher}$, and $\alpha_{\rm ELBM}$ at time $t= 500$ for viscosity $\nu=1.0 \times 10^{-5}$. }
    \label{shock_tube}
\end{figure*}      
 
To compare the behaviour of $\alpha_{\rm Lower},\alpha_{\rm Higher}$ with $\alpha_{\rm ELBM}$ and $\alpha_{\rm LBGK}$, we first simulate the one-dimensional shock tube using the $D1Q3$ lattice. 
In this setup, a domain with 800 grid points is initialized with a step function for density  as  $\rho\,(x \leq 400) = 1.5$ and $\rho\,(x>400) = 0.75$.
The presence of a sharp discontinuity in the initial condition at the center of the domain generates a moving compressive shock front in the low-density region and a rarefaction front in the high-density region. 
These two fronts give rise to a contact region of uniform pressure and velocity in the center of the tube \citep{laneyCGD}. 
The density, velocity, and entropy profiles shown in Figure \ref{shock_tube} illustrate that the numerical oscillations are sharply reduced in the case of $\alpha_{\rm Lower}$, thus pointing to its dissipative nature. 
It can also be seen that the oscillations are prominent for $\alpha_{\rm LBGK}$ and that both $\alpha_{\rm Higher}$ and $\alpha_{\rm ELBM}$ restore the $H$ theorem without altering the fields.

\begin{figure}
   \centering
    {\includegraphics[width=0.4\textwidth]{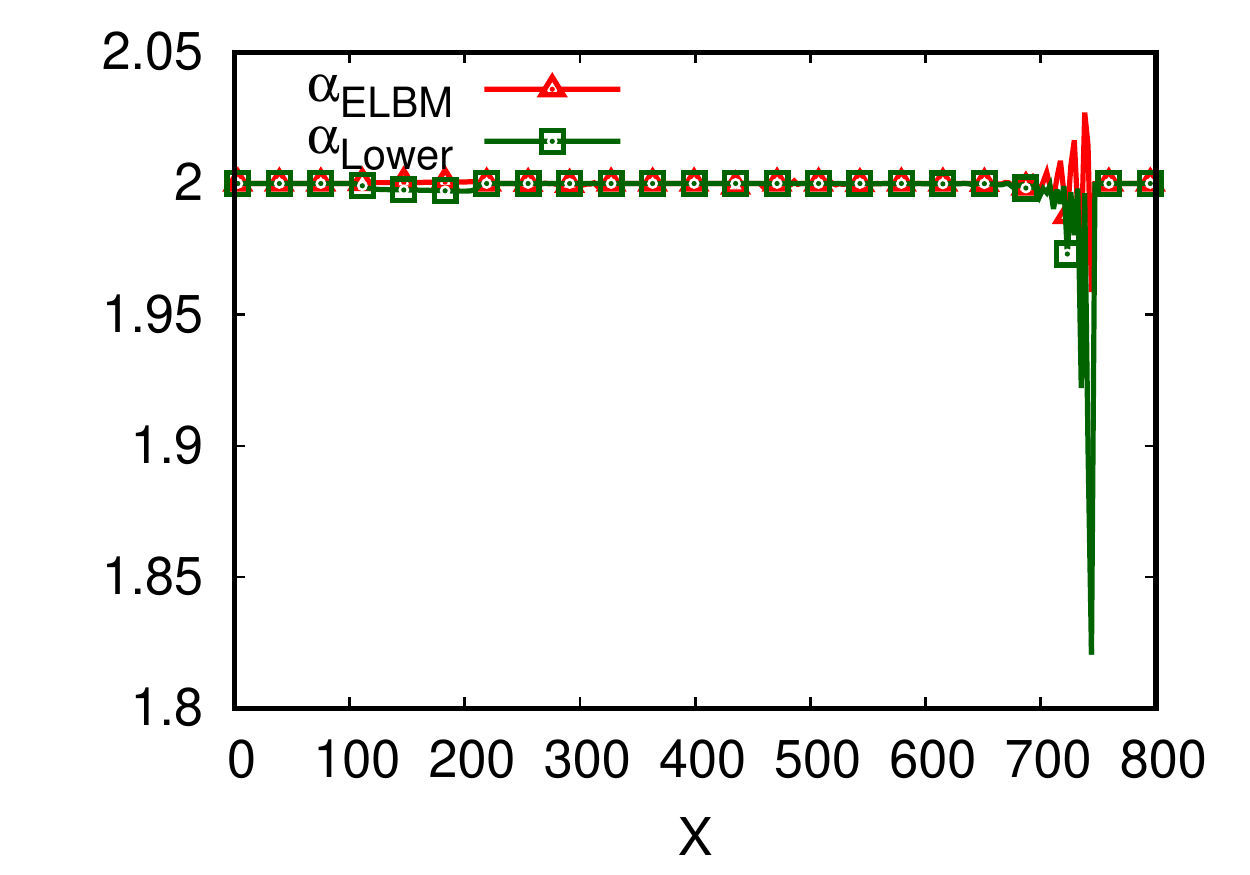}}\hspace{1cm}
    {\includegraphics[width=0.4\textwidth]{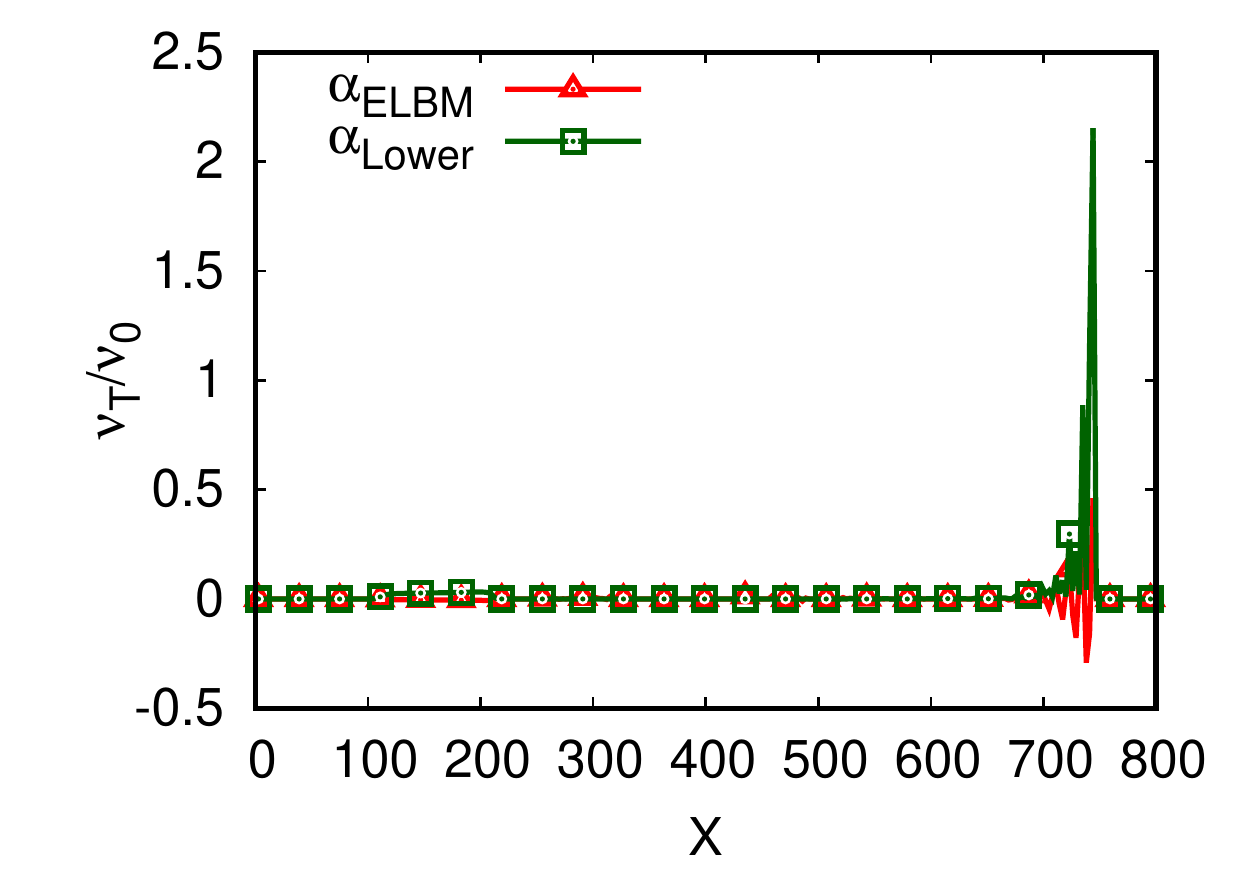}} 
    \caption{Comparison between $\alpha_{\rm Lower}$ and $\alpha_{\rm ELBM}$ for the Sod shock tube. Top: Snapshot of the path length.
    Bottom: Ratio of the turbulent viscosity correction to the kinematic viscosity. }
    \label{alpha}
\end{figure}

\begin{figure}
   \centering
    {\includegraphics[width=0.4\textwidth]{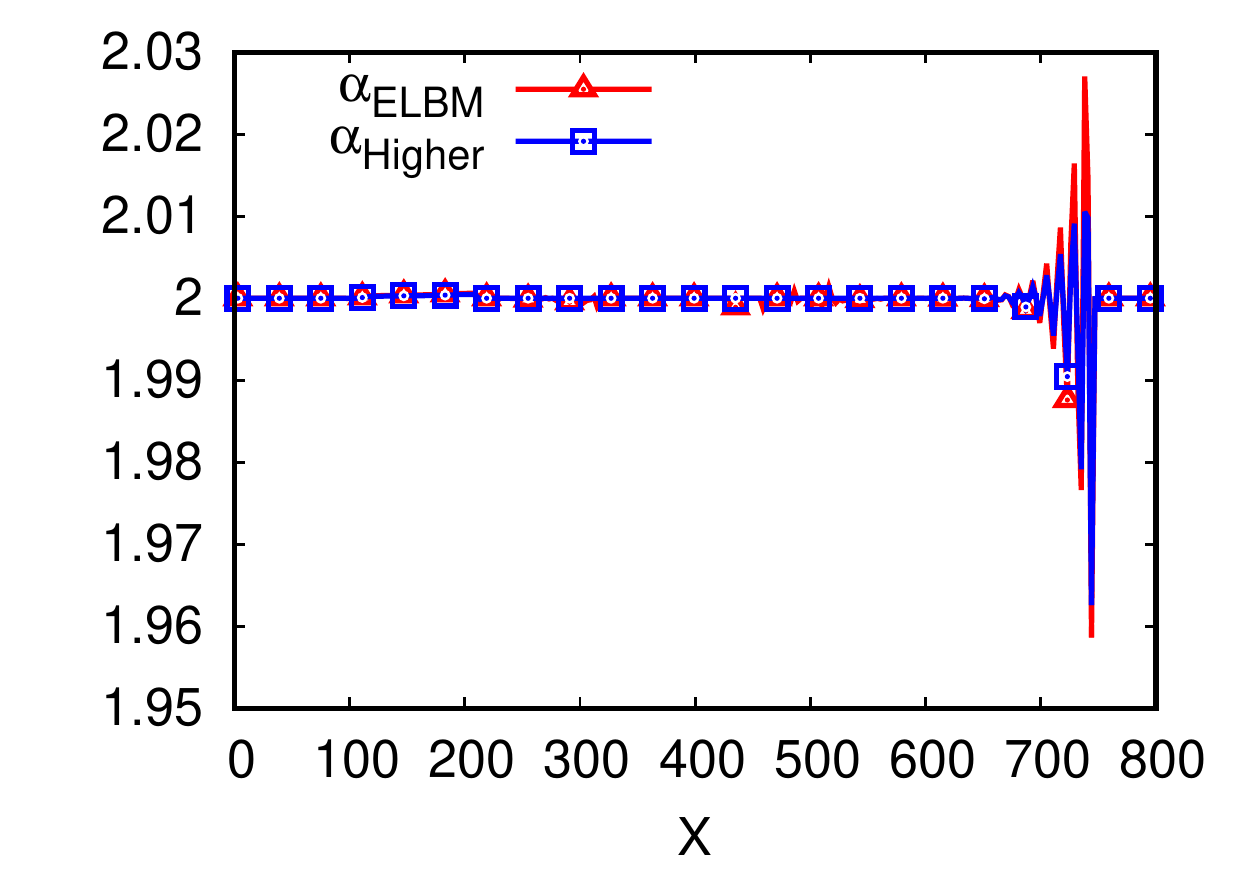}}\hspace{1cm}
    {\includegraphics[width=0.4\textwidth]{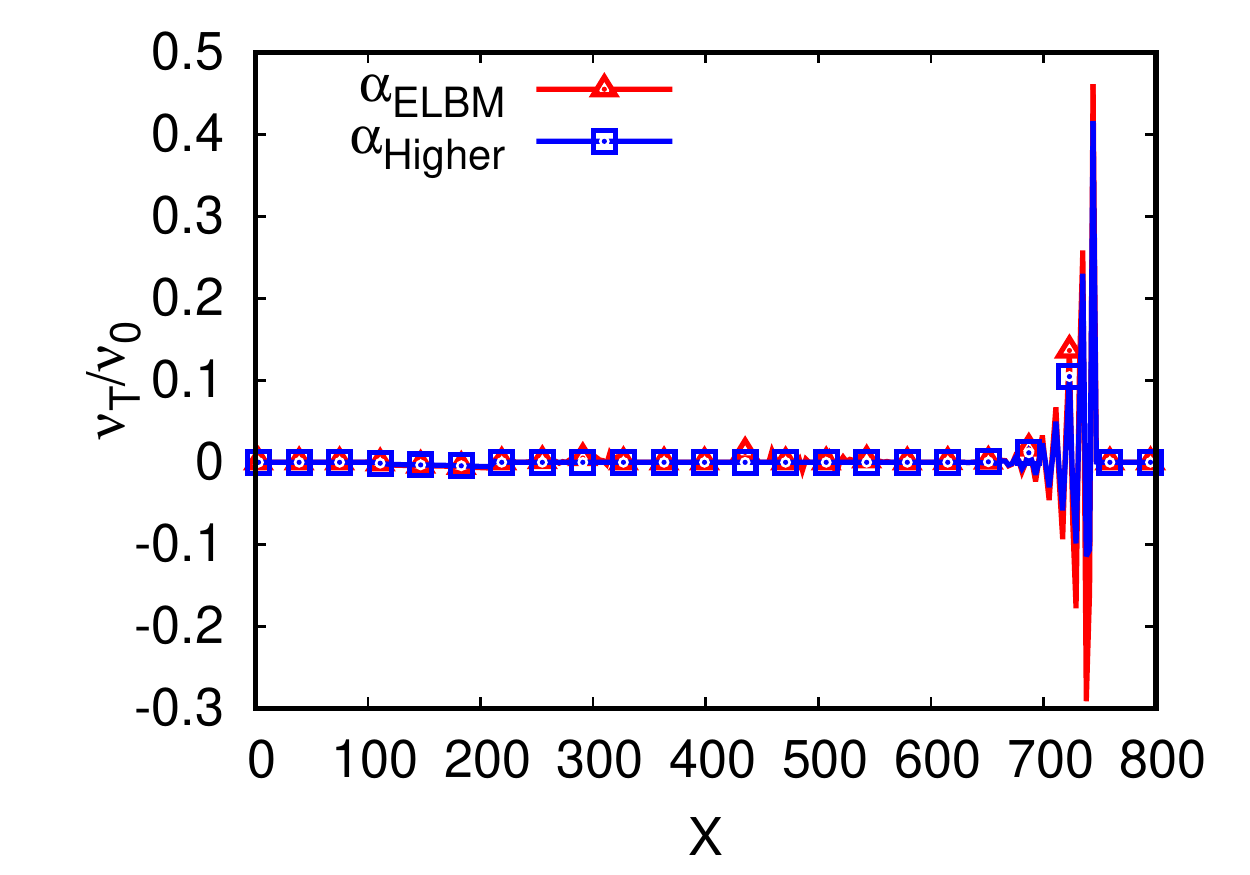}} 
    \caption{Comparison between $\alpha_{\rm Higher}$ and $\alpha_{\rm ELBM}$ for the Sod shock tube. Top: Snapshot of the path length.  Bottom: Ratio of the turbulent viscosity correction to the kinematic viscosity. }
    \label{alphaH}
\end{figure}

Figure \ref{alpha} (top) compares $\alpha_{\rm Lower}$ and $\alpha_{\rm ELBM}$. It is evident that the path lengths show departure from $\alpha=2$ (BGK value) only in the narrow regions of the compressive and the rarefaction fronts. 
It can also be seen that the value of $\alpha_{\rm Lower}$ is always smaller than 2, while that of $\alpha_{\rm ELBM}$ fluctuates about 2.  
Figure \ref{alpha} (bottom) plots the ratio of turbulent viscosity correction to kinematic viscosity $\nu_T/\nu_0$ (more details in Sec. \ref{subvisc}). From the figure, it is evident that at the location of the shock front the $\alpha_{\rm Lower}$ is more than twice the kinematic viscosity, while $\alpha_{\rm ELBM}$ is only $\sim 47\%$. 
Similarly, Figure \ref{alphaH} (top) compares the path length from $\alpha_{\rm Higher}$ and it is seen that for this setup $\alpha_{\rm Higher}$ exhibits smaller fluctuations than the $\alpha_{\rm ELBM}$. Figure \ref{alphaH} (bottom) shows that the turbulent viscosity correction for $\alpha_{\rm ELBM}$ is $\sim 47\%$, whereas for $\alpha_{\rm Higher}$ it is $\sim 42\%$.
Hence, it can be concluded that $\alpha_{\rm Higher}$ imposes the $H$ theorem (thus guaranteeing unconditional numerical stability) with the least turbulent viscosity correction.
% In Sec. \ref{subvisc}, we demonstrate that departure of $\alpha$ from 2 can be interpreted as a turbulent viscosity correction $\nu_T$. 

\subsection{Doubly periodic shear layer}

In this section, we compare the behaviour of $\alpha_{\rm Lower}, \alpha_{\rm Higher}$ with $\alpha_{\rm LBGK}$ by considering the setup of doubly periodic shear layer \citep{minion1997performance}. 
The initial velocity field comprises of two shear layers given by
\begin{align}
u_x (y) &= 
\begin{cases}
    U_0 \tanh [ (4y-1)/w], \qquad y \leq 1/2 \\
    U_0 \tanh [ (3-4y)/w], \qquad y>1/2               
\end{cases}\\
u_y (x) &= U_0 \delta \sin [ 2\pi (x+1/4) ],
\end{align}
where $w=\delta=0.05, U_0=0.04$ and $x,y$ are nondimensionalized coordinates. The viscosity is calculated from the Reynolds number which for the present case is fixed at $3\times 10^4.$ 
It is known that at poor grid resolutions for this setup, the numerical disturbances may lead to formation of spurious vortices in the braids \citep{minion1997performance,coreixas2017recursive}.

\begin{figure}\centering
{\includegraphics[scale=0.25]{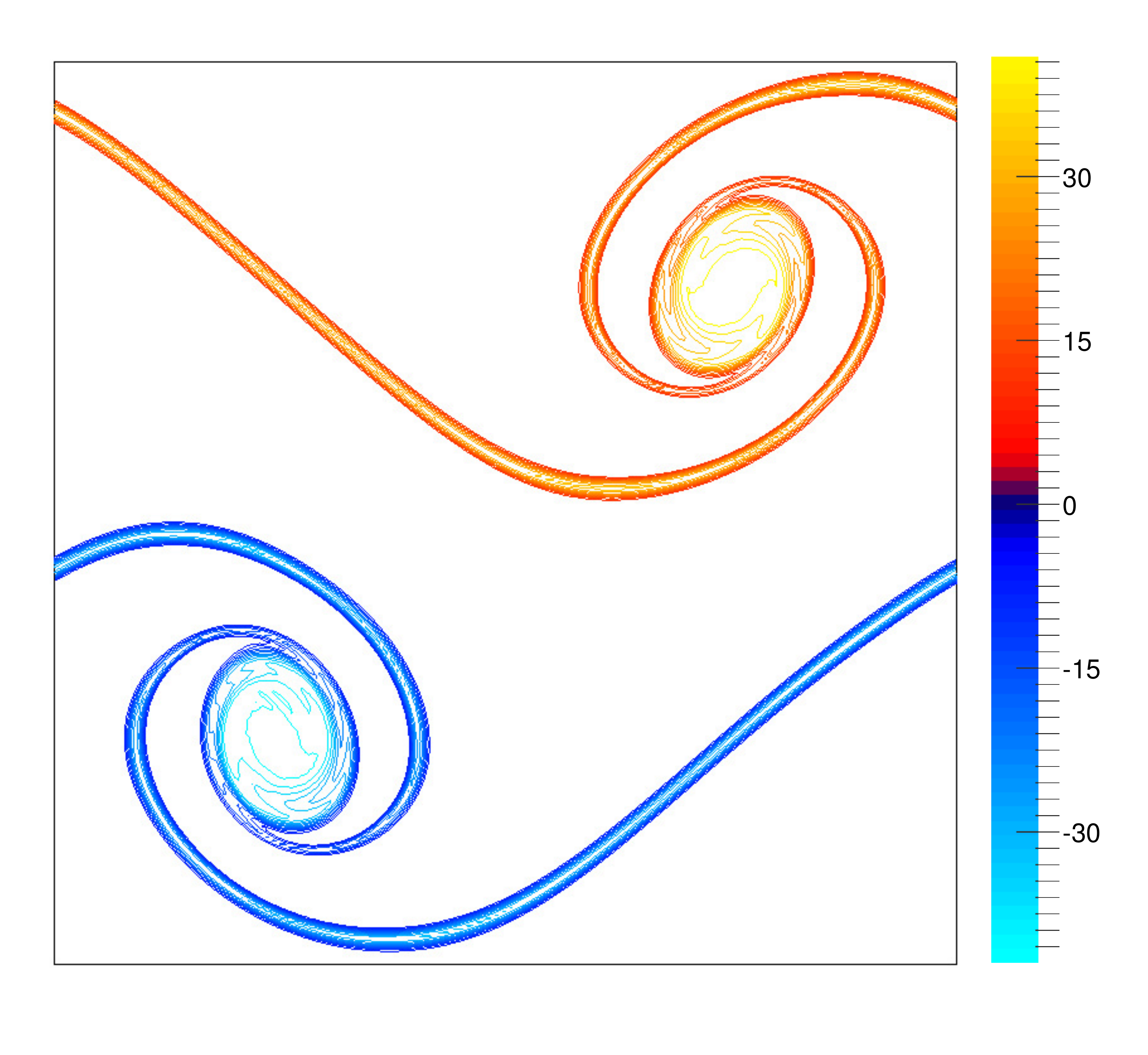} }
{\includegraphics[scale=0.25]{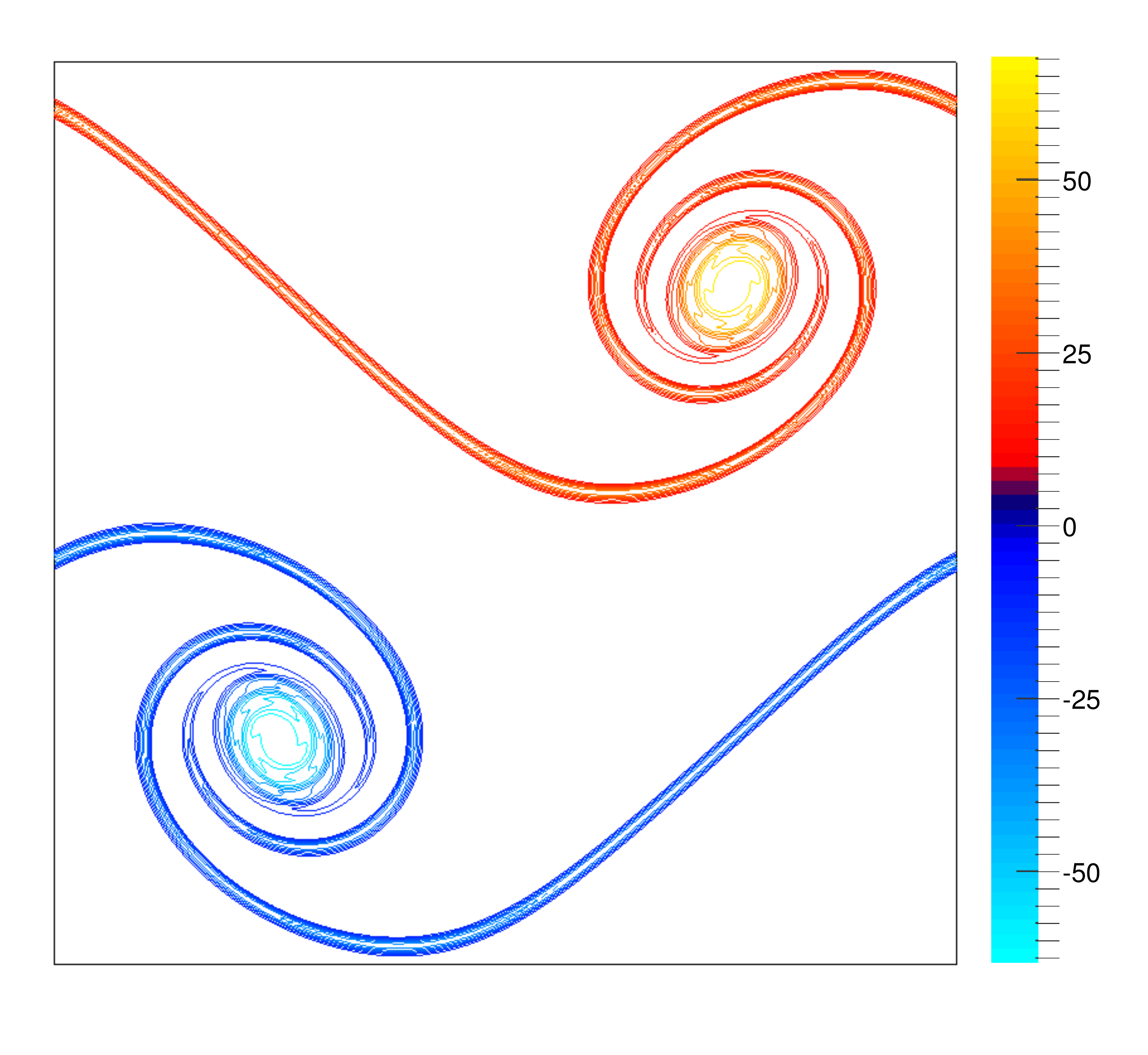} }
{\includegraphics[scale=0.25]{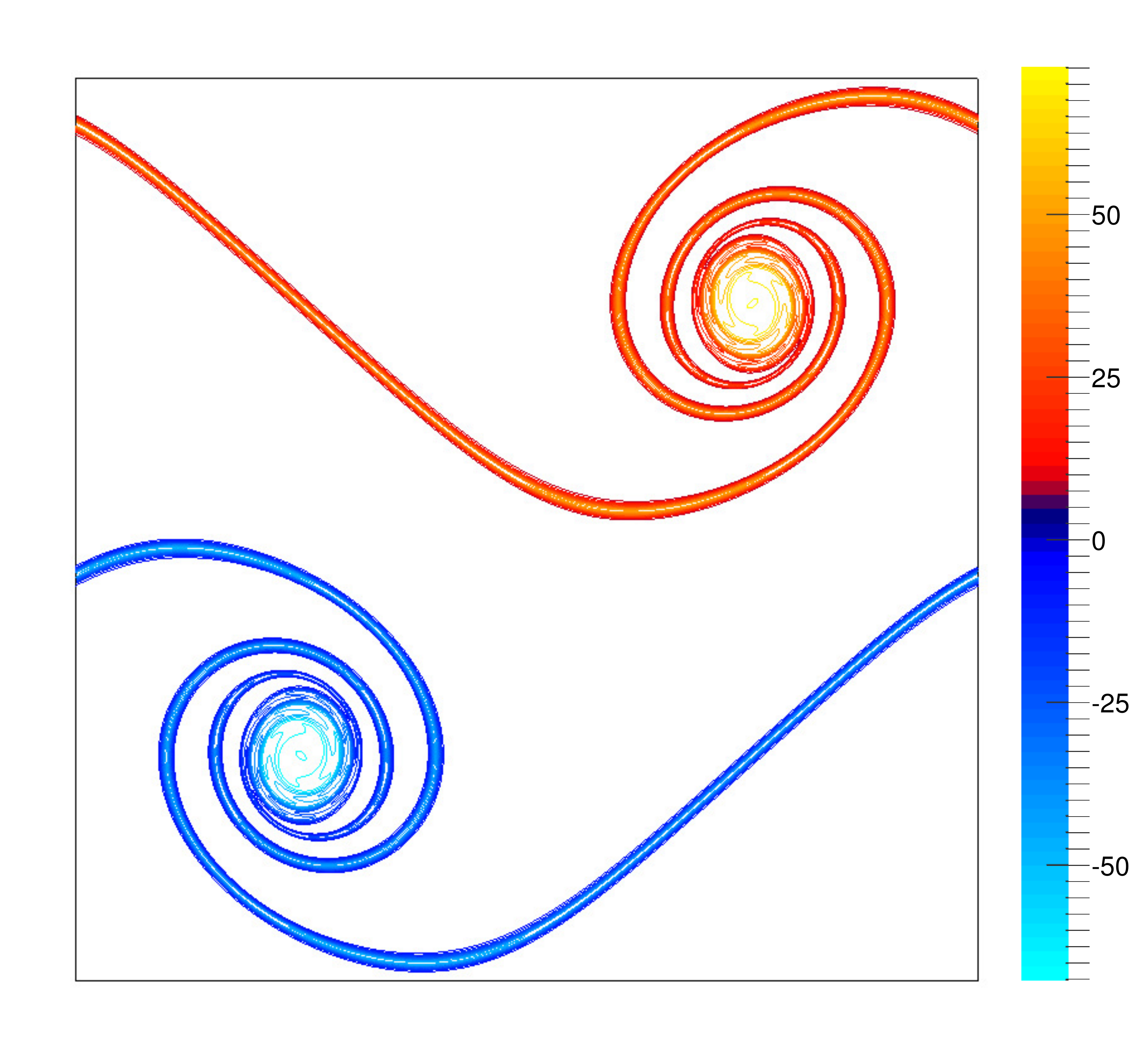} }
\caption{Nondimensional iso-vorticity contours for $\alpha_{\rm Lower}$ (left), $\alpha_{\rm Higher}$ (center) at grid size $256\times 256$ and for BGK at $1024\times 1024$ (right) after one convection time.}
\label{dpsl_vorticity}
\end{figure}

\begin{figure*}
{\includegraphics[trim={0em 0 0 0},clip,scale=0.25]{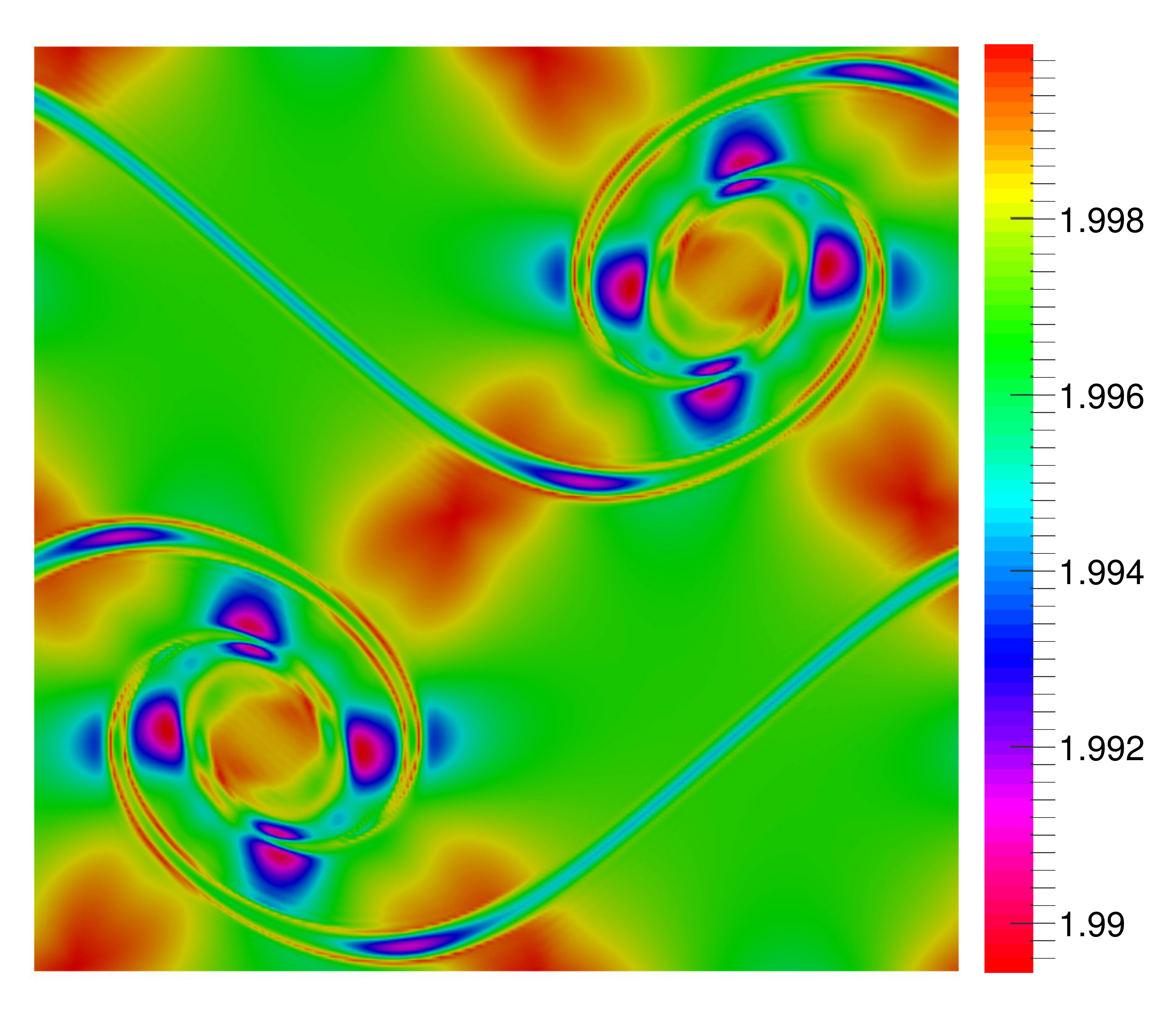} }
{\includegraphics[trim={8em 0 0 0},clip,scale=0.25]{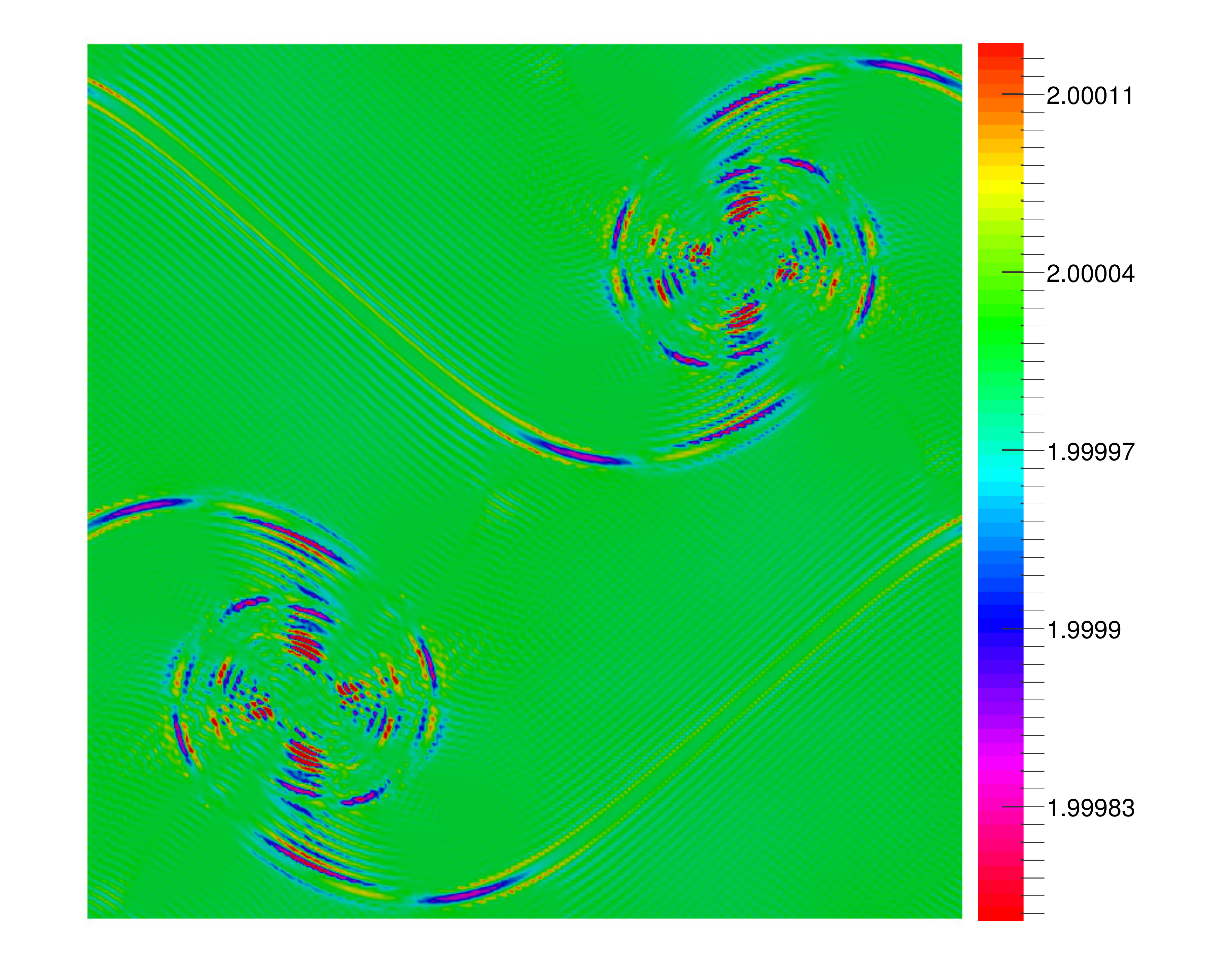} }
\caption{Path length from $\alpha_{\rm Lower}$ (left) and $\alpha_{\rm Higher}$ (right) after one convection time on a grid of size $256\times 256$.}
\label{dpsl_alpha}
\end{figure*}

\begin{figure}\centering
{\includegraphics[scale=0.7]{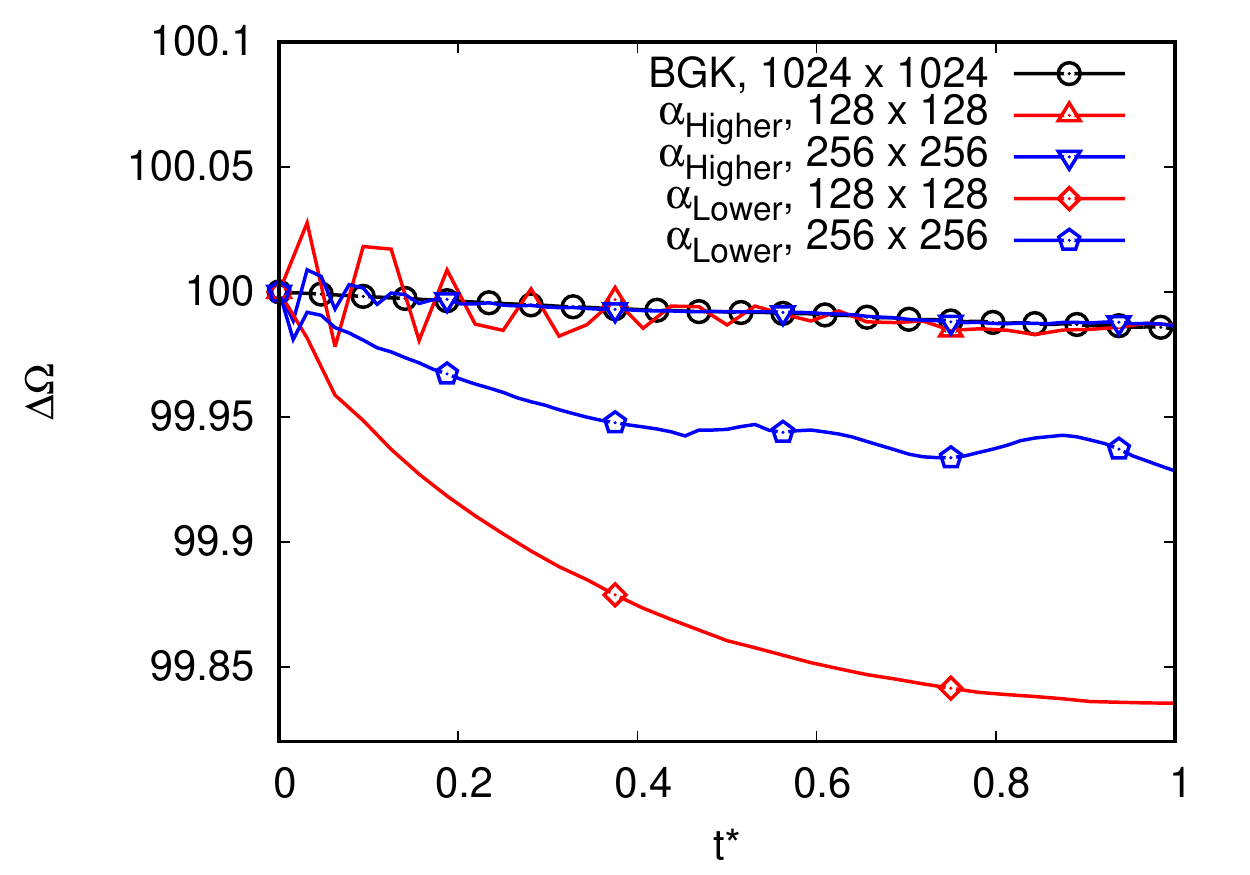} }
\caption{Change in the global enstrophy $\Delta \Omega$ vs time for various square grids.  Here, $t^*$ is the nondimensional convection time.}
\label{dpsl_enstrophy}
\end{figure}

Figure \ref{dpsl_vorticity} depicts the isovorticity contours for $\alpha_{\rm Lower}, \alpha_{\rm Higher}$ on a $256\times 256$ grid and for $\alpha_{\rm LBGK}$ on $1024\times 1024$ grid obtained after one convection time.
A qualitative comparison of the three plots reveal that the vortex structure is smudged for $\alpha_{\rm Lower}$, while the vortex structure of $\alpha_{\rm Higher}$ on a $256\times 256$ grid
is the same as that of BGK at $1024\times 1024$ grid.
In Fig. \ref{dpsl_alpha} we show the magnitude of the path lengths $\alpha_{\rm Lower}, \alpha_{\rm Higher}$, from where it evident that while $\alpha_{\rm Lower}$ always remains smaller than 2, $\alpha_{\rm Higher}$ fluctuates about 2, thus corroborating the dissipative nature of $\alpha_{\rm Lower}$.
Finally, a quantitative analysis of the flow is performed by measuring the change in global enstrophy ($\Delta\Omega = \bar\Omega_t/\bar\Omega_{0} \times 100$), where $\bar\Omega_t$ is the enstrophy at time $t$, and $\bar\Omega_{0}$ is the initial global enstrophy (defined as the square of the vorticity).
Figure \ref{dpsl_enstrophy} plots the time evolution of $\Delta\Omega $. 
It is evident that $\alpha_{\rm Higher}$ on a $128 \times 128$ grid behaves the same as the BGK on a much larger $1024\times 1024$ grid, whereas $\alpha_{\rm Lower}$ exhibits dissipation that manifests in the form of reduced enstrophy.

\subsection{Lid-driven cavity}

In this section, we consider the lid-driven cavity at a Reynolds number ($\rm Re$) of 5000 where the motion of the top wall drives the flow in a 2D cavity.
We use the standard $D2Q9$ lattice and diffuse boundary condition \cite{ansumali2002kinetic}. 
For this setup, the LBGK ($\alpha=2$) is numerically unstable at smaller grid sizes of $64\times 64, \, 96\times 96$, and $128\times 128$, however, it is stable at a larger grid of size $256\times 256$. 
The entropic formulations $\alpha_{\rm Lower},\alpha_{\rm Higher},\alpha_{\rm ELBM}$ are stable at all grid sizes. 

\begin{figure*}
   \centering
    {\includegraphics[trim={6.cm 1cm 6cm 1cm},clip,width=0.3\textwidth]{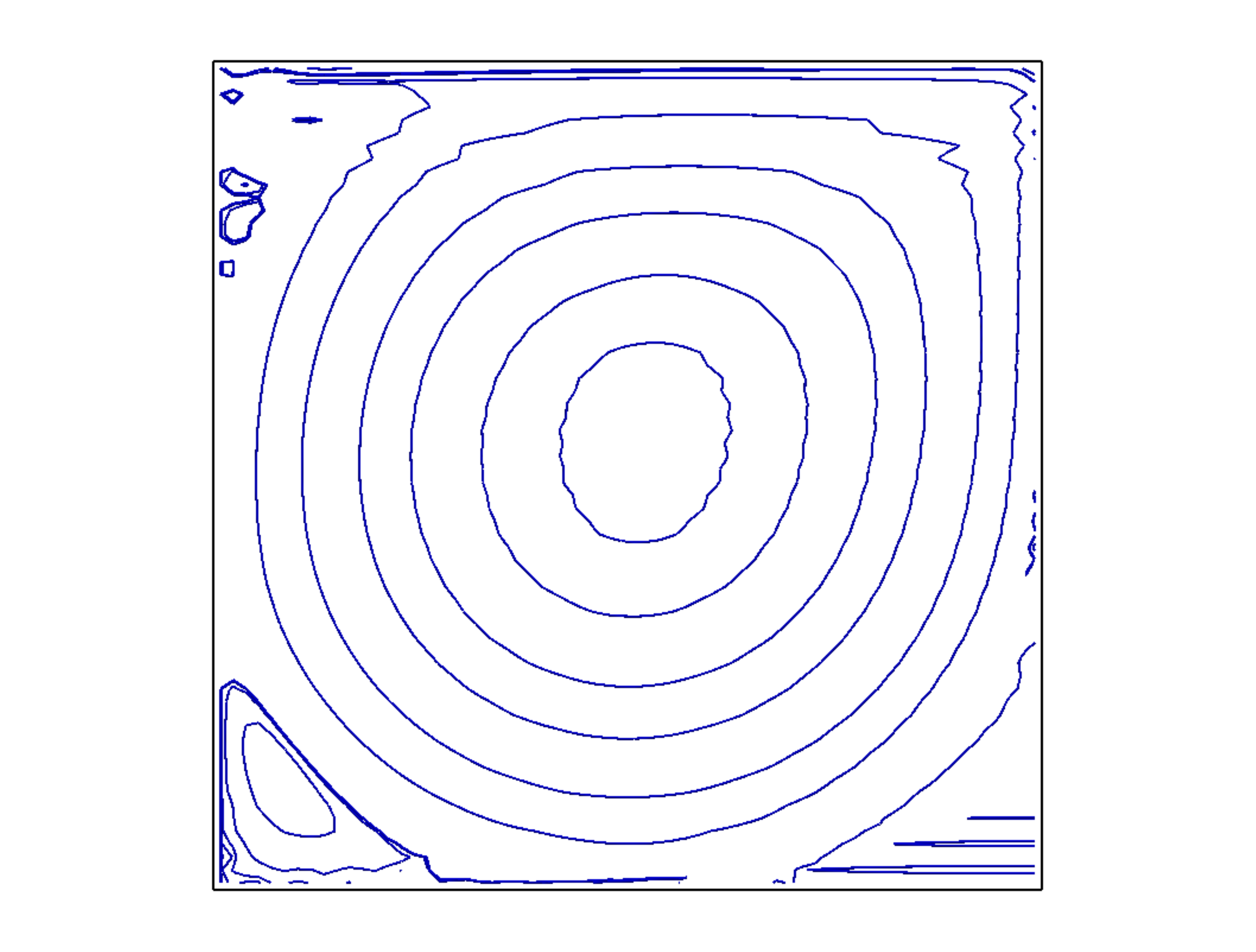}}
    {\includegraphics[trim={6cm 1cm 6cm 1cm},clip,width=0.3\textwidth]{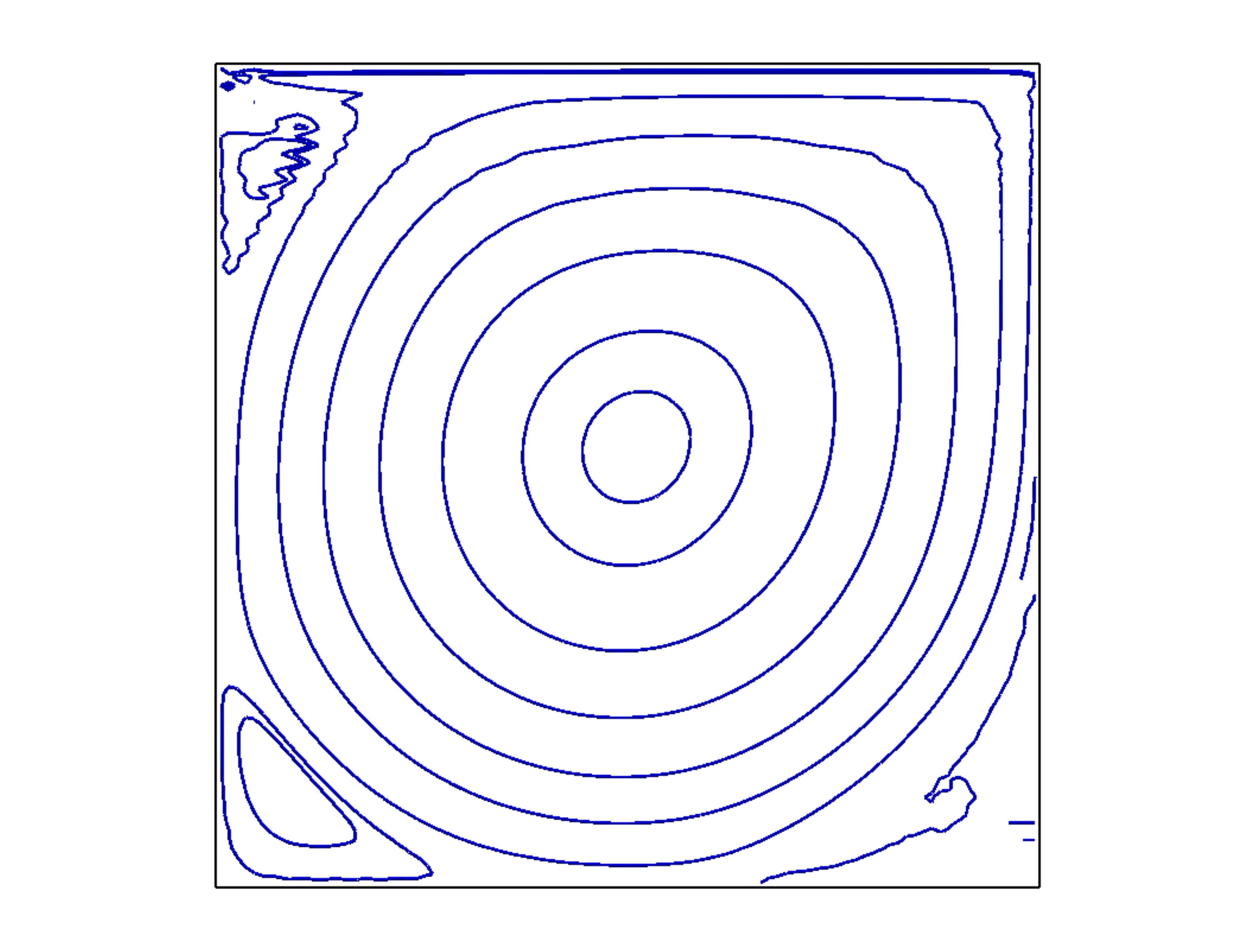}}
    {\includegraphics[width=0.307\textwidth]{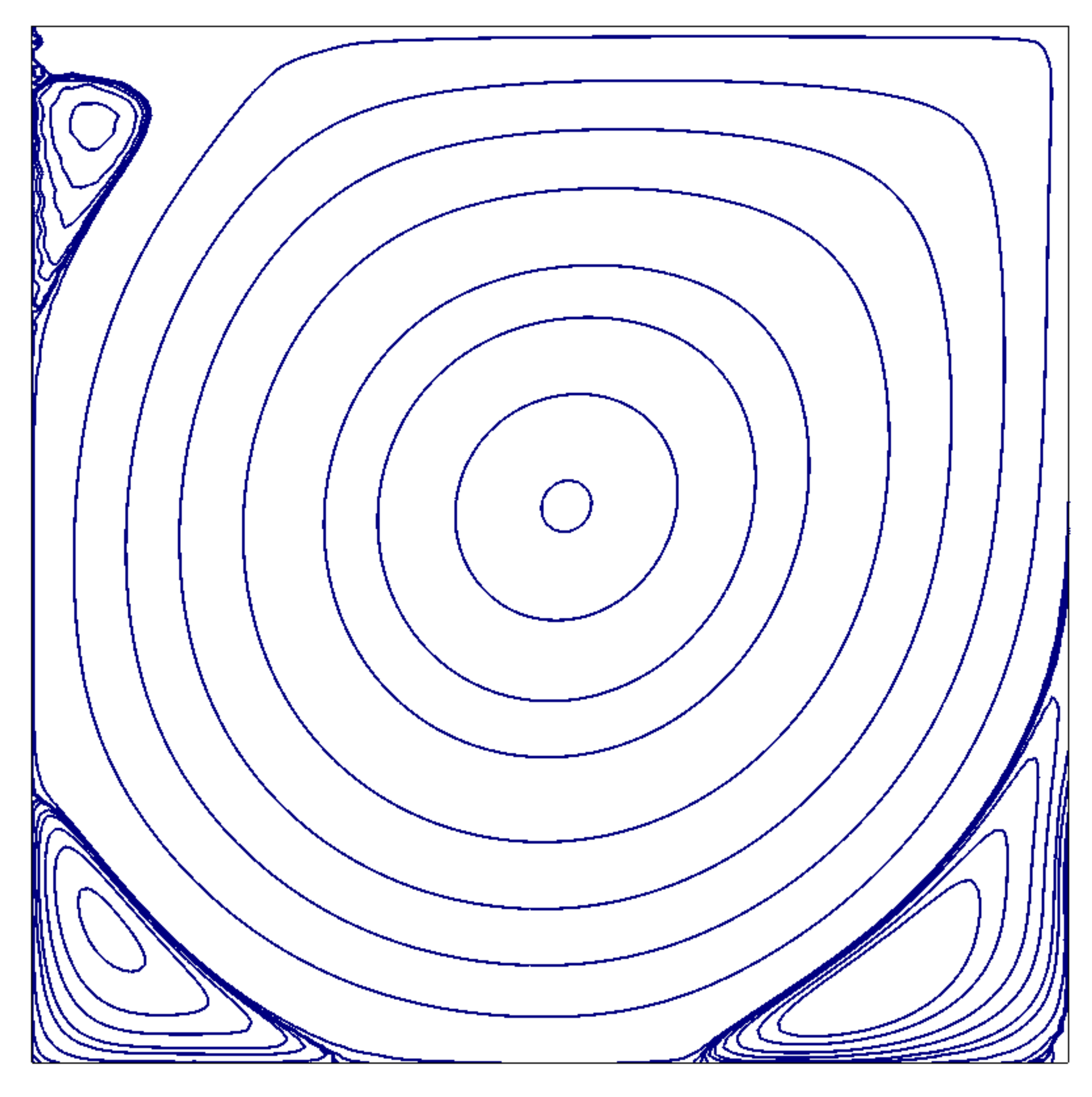}} 
    \caption{Iso-vorticity contours for the lid-driven cavity at Reynolds number of 5000 for various grid sizes: $64\times 64$ (left), $96\times 96$ (center), $128\times 128$ (right). }
    \label{isovorticityldc}
\end{figure*}

\begin{figure}
   \centering
    {\includegraphics[width=0.45\textwidth]{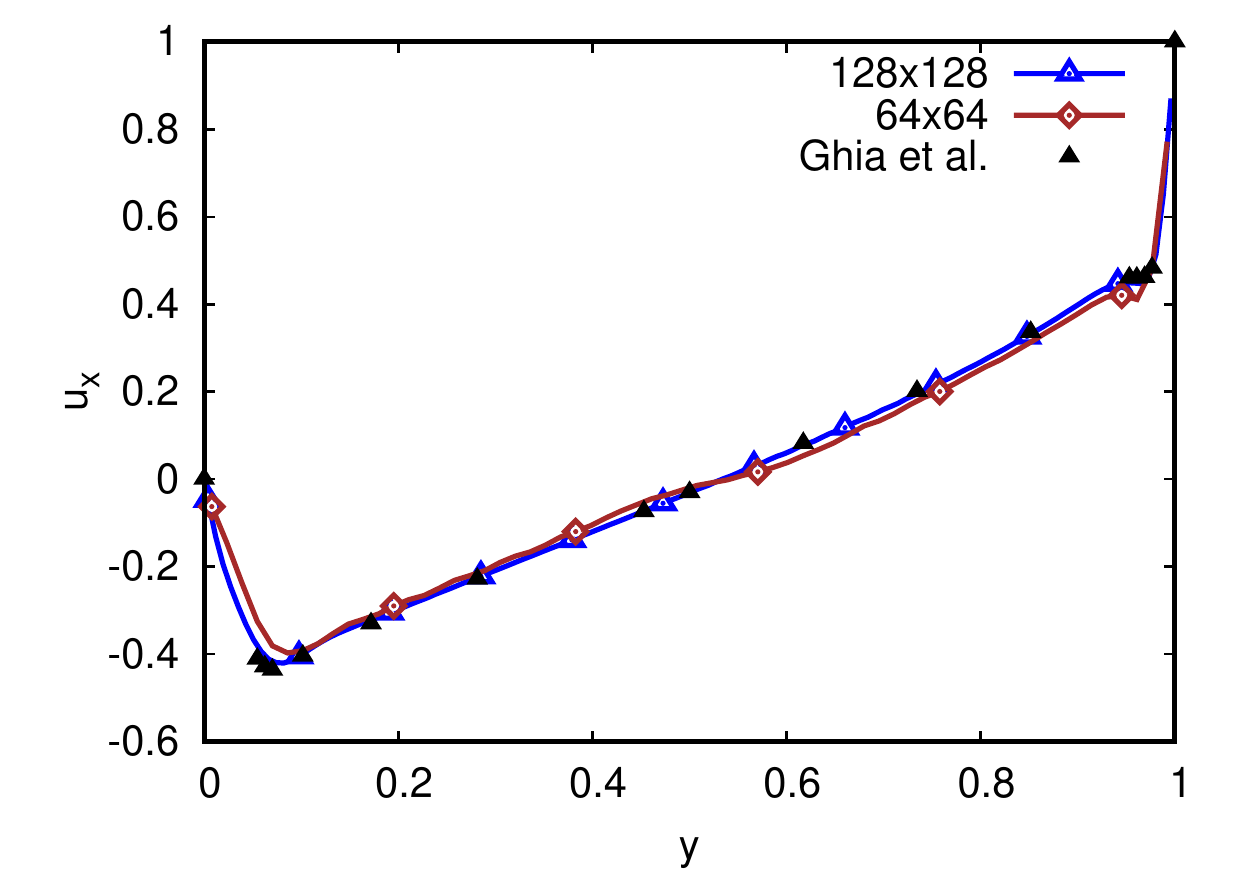}}
    {\includegraphics[width=0.45\textwidth]{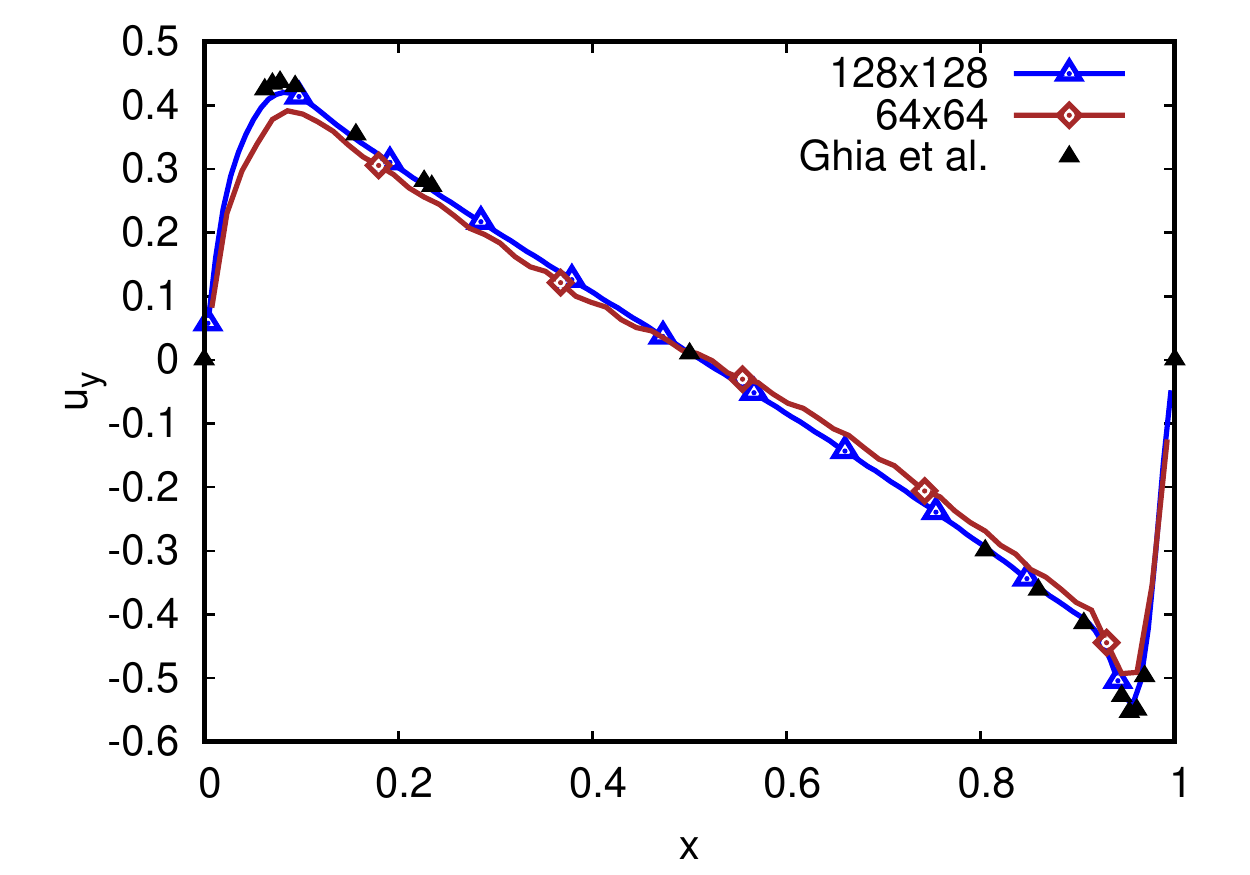}}
    \caption{Velocity profiles for the lid-driven cavity at Reynolds number of 5000 and Mach number 0.05 for various grid sizes. Top: nondimensionalized x-velocity along the vertical centerline. Bottom: nondimensionalized y-velocity along the horizontal centerline.  }
    \label{velocityldc}
\end{figure}

Figure \ref{isovorticityldc} depicts the iso-vorticity contours for various grid sizes obtained using $\alpha_{\rm Higher}$. It is seen that even extremely under-resolved grids remain numerically stable. However, at coarse resolutions like $64\times 64$ and $96\times 96$ the finer structures are distorted, which take the expected form at a slightly higher grid size of $128\times 128$. It should be repeated here that at grid size of  $128\times 128$ the LBGK ($\alpha=2$) is numerically unstable. In Fig. \ref{velocityldc}, we plot the velocities along vertical and horizontal centerlines and observe a good match with \cite{ghiaghia}. 

% We also compare the the time evolution of enstrophy obtained using $\alpha_{\rm LBGK}$ at grid size $256\times 256$ with that of $\alpha_{\rm Lower}, \alpha_{\rm Higher},\alpha_{\rm ELBM}$ at $96\times 96$ and $128\times 128$ in Fig. \ref{ldchigher}. 
% It is evident that the enstrophy predicted by $\alpha_{\rm ELBM},\alpha_{\rm Higher}$ and $\alpha_{\rm LBGK}$ are comparable, while that of $\alpha_{\rm Lower}$ is smaller.
% Therefore, we can conclude that the solution provided by $\alpha_{\rm Lower}$ is dissipative. 
% This is attributed to the high magnitude of $ H^{(B)}$ in Eq. \eqref{compactLower} due to loose bounds on the logarithms.
% 
% \begin{figure}
%    \centering
%     {\includegraphics[width=0.35\textwidth]{./figures/eelbmComparisonEnstrophy96}}
%     {\includegraphics[width=0.35\textwidth]{./figures/eelbmComparisonEnstrophy128}} 
%     \caption{Comparison of time evolution of global enstrophy $\bar \Omega$ between $\alpha_{\rm Lower},\alpha_{\rm Higher},\alpha_{\rm ELBM}$ for the lid-driven cavity at ${\rm Re}=5000$ and ${\rm Ma}=0.05$. The BGK solution is at a grid size of $256\times 256$, however,  $\alpha_{\rm Lower},\alpha_{\rm ELBM}$ are at $96\times 96$ (left) and $128\times 128$ (right). Here, $t^*$ is time non-dimensionalized via the convective time scale. The dissipative nature of $\alpha_{\rm Lower}$ manifests in the form of reduced enstrophy.}
%     \label{ldchigher}
% \end{figure}

\begin{figure}
   \centering
    {\includegraphics[width=0.5\textwidth]{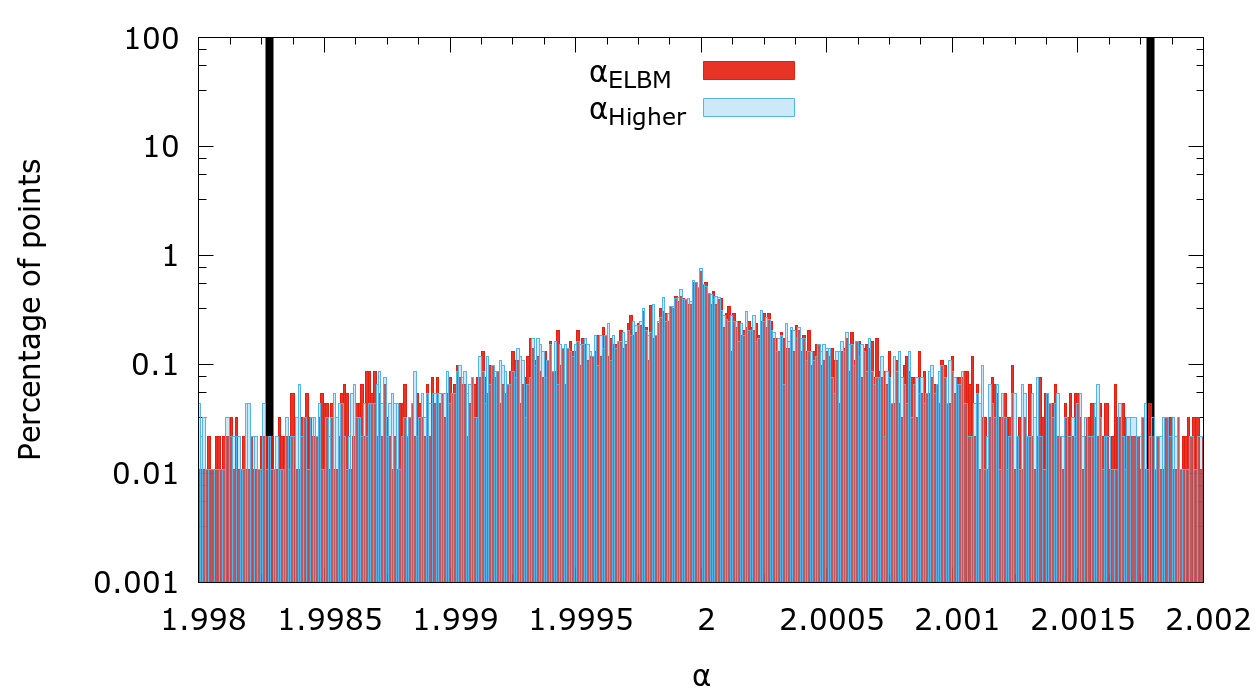}} 
    {\includegraphics[width=0.5\textwidth]{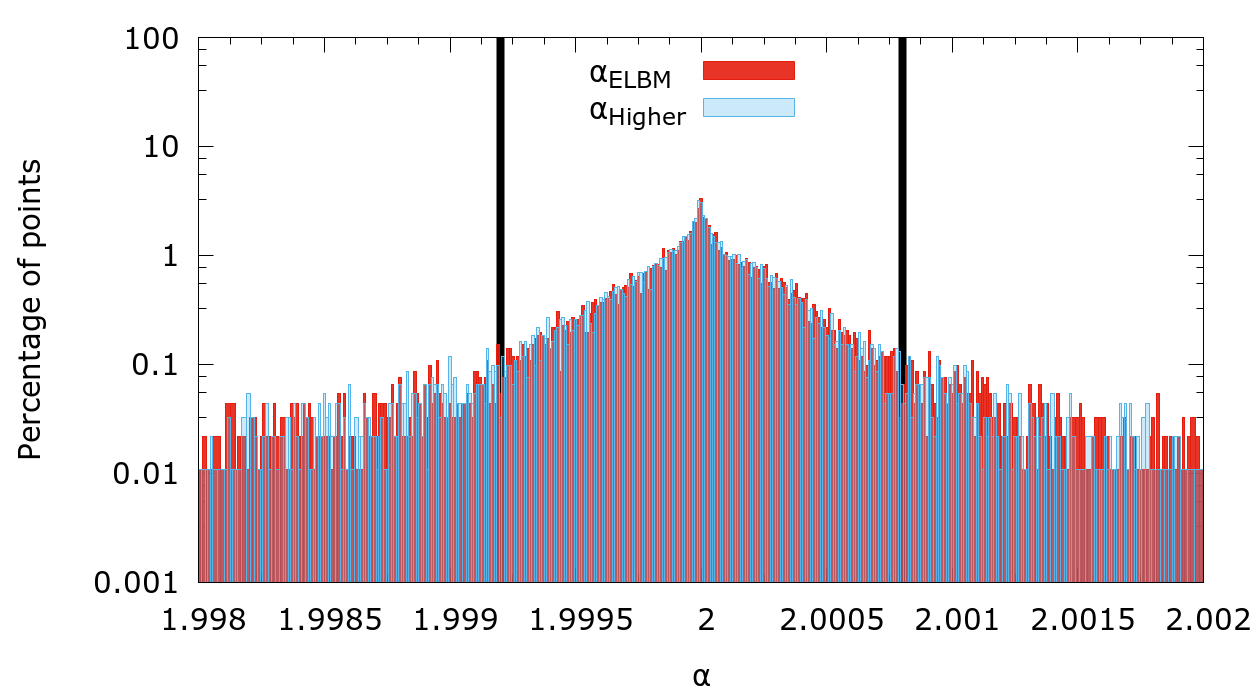}}\\
    {\includegraphics[width=0.5\textwidth]{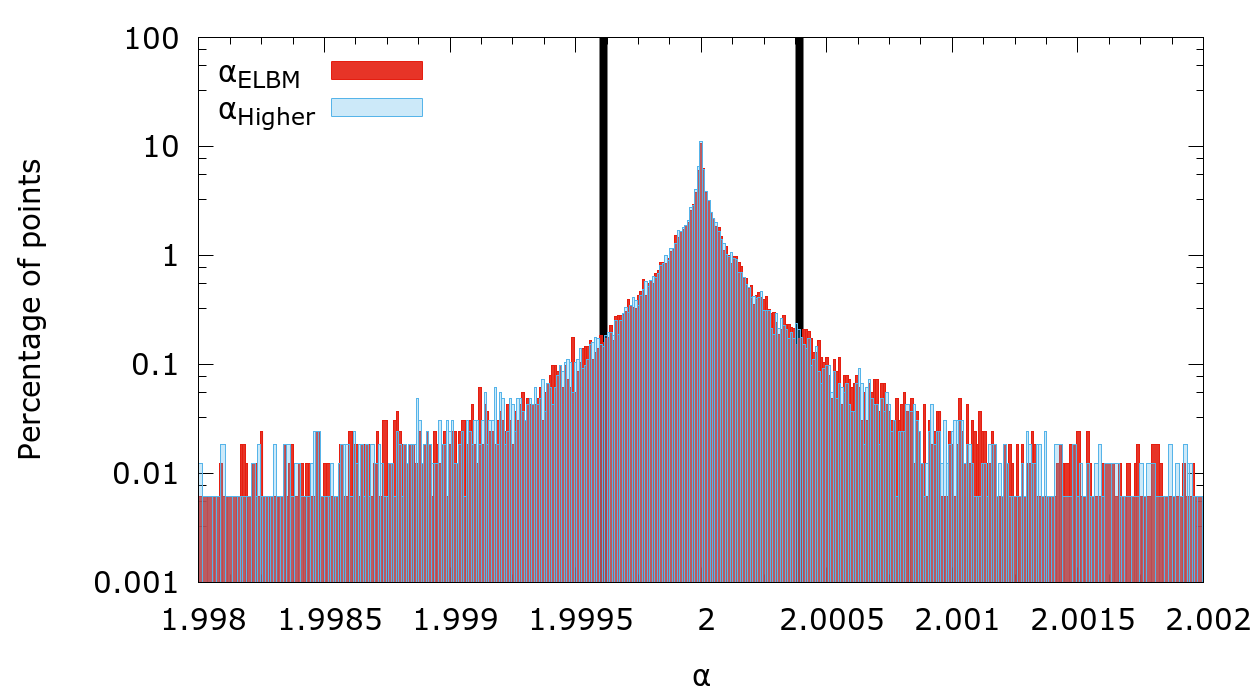}} 
    \caption{Distribution of $\alpha_{\rm Higher}$ and $ \alpha_{\rm ELBM}$  for lid-driven cavity at Reynolds number 5000 and Mach number 0.05. Grid sizes are $64\times 64$ (top), $96\times 96$ (middle), $128\times 128$ (bottom). The difference between the distribution of $\alpha_{\rm Higher}$ and $\alpha_{\rm ELBM}$ is seen to be insignificant. The solid black lines denote the region inside which $90\%$ of the points lie. The locations of the solid lines are tabulated in Table \ref{alpha90tab}.}
    \label{ldcalphadistsame}
\end{figure}

Next, we establish that there is no appreciable difference between the path lengths $\alpha_{\rm Higher}$ and $\alpha_{\rm ELBM}$. 
To this effect, we compare the instantaneous value of $\alpha_{\rm Higher}$ and $\alpha_{\rm ELBM}$ for three different grid resolutions. 
First, the simulation is performed using $\alpha_{\rm Higher}$ for 100 convection times.
On the populations thus obtained, we evaluate $\alpha_{\rm Higher}$ and $\alpha_{\rm ELBM}$ for the entire grid.
The $L_1, \, L_2, \, L_{\infty}$ error norms of $||\alpha_{\rm Higher}-\alpha_{\rm ELBM}||$ are tabulated in Table \ref{alphaHighertab}, whereas the distribution of path lengths are given in Fig. \ref{ldcalphadistsame}. 
It is evident that $\alpha_{\rm Higher}$ and $\alpha_{\rm ELBM}$ show insignificant deviation at all grid sizes. 
From  Fig. \ref{ldcalphadistsame} and Table \ref{alpha90tab}, it can also be seen that as the grid size increases the distribution of the path lengths becomes narrower as the region around the LBGK value of $\alpha=2$ where $90\%$ of the points lie (inside solid vertical lines) becomes smaller.

\begin{table}\centering
\begin{tabular}{|c|c|c|c|}
\hline
             & $64\times 64$        & $96\times 96$  & $128\times 128$ \\\hline
$L_1$        & $2.55\times 10^{-5}$ & $1.37\times 10^{-5}$ & $8.26\times 10^{-6}$ \\\hline
$L_2$        & $1.67\times 10^{-4}$ & $9.89\times 10^{-5}$ & $5.17\times 10^{-5}$ \\\hline
$L_{\infty}$ & $6.07\times 10^{-3}$ & $5.24\times 10^{-3}$ & $3.71\times 10^{-3}$ \\\hline
\end{tabular}
\caption{Error norms for $||\alpha_{\rm Higher} - \alpha_{\rm ELBM}||$.}
\label{alphaHighertab}
\end{table}

\begin{table}\centering
\begin{tabular}{|c|c|c|c|}
\hline
      & $64\times 64$ & $96\times 96$   & $128\times 128$ \\\hline
$\alpha_{\rm Higher}$ & $2\pm 1.79\times 10^{-3}$ & $2\pm 8.0\times 10^{-4}$ & $2\pm 3.9\times 10^{-4}$ \\\hline
$\alpha_{\rm ELBM}$   & $2\pm 1.77\times 10^{-3}$ & $2\pm 7.3\times 10^{-4}$ & $2\pm 2.9\times 10^{-4}$ \\\hline
\end{tabular}
\caption{Region around the LBGK value of $\alpha=2$ where $90\%$ of the points lie. It is seen that as the grid size increases the region becomes narrower.}
\label{alpha90tab}
\end{table}

% \begin{figure}
%    \centering
%     {\includegraphics[width=0.32\textwidth]{./figures/alphaHigherMinusKarlin64}}
%     {\includegraphics[width=0.32\textwidth]{./figures/alphaHigherMinusKarlin96}} 
%     {\includegraphics[width=0.32\textwidth]{./figures/alphaHigherMinusKarlin128}}
%     \caption{Distribution of $\alpha_{\rm Higher} - \alpha_{\rm ELBM}$  for lid-driven cavity at Reynolds number 5000 and Mach number 0.05. Grid sizes are $64\times 64$ (left), $96\times 96$ (center), $128\times 128$ (right). The difference between the distribution reduces as the grid size increases. }
%     \label{ldcalphasnapsame}
% \end{figure}

We also briefly investigate the idea that the path length $\alpha_{\rm Lower}$ could be utilized as a good initial guess value for the iterative ELBM solver.
Typically, the iterative root solver converges in 4-5 iterations, however, it is stipulated that the converged result should be obtained in a single iteration when using $\alpha_{\rm Lower}$ as the initial guess value.
We call this first iterate $\alpha_{\rm Iterate1}$ and compare it with $\alpha_{\rm ELBM}$.
The $L_1, L_2, L_{\infty}$ error norms of $||\alpha_{\rm Iterate1}-\alpha_{\rm ELBM}||$ are tabulated in Table \ref{alphaIter1tab} from where it can be concluded that the difference is insignificant for all three grid sizes.

\begin{table}\centering
\begin{tabular}{|c|c|c|c|}
\hline
             & $64\times 64$        & $96\times 96$        & $128\times 128$ \\\hline
$L_1$        & $2.27\times 10^{-7}$ & $7.48\times 10^{-8}$ & $2.56\times 10^{-8}$ \\\hline
$L_2$        & $3.02\times 10^{-6}$ & $1.37\times 10^{-6}$ & $8.26\times 10^{-7}$ \\\hline
$L_{\infty}$ & $1.10\times 10^{-4}$ & $9.00\times 10^{-5}$ & $7.00\times 10^{-5}$ \\\hline
\end{tabular}
\caption{Error norms for $||\alpha_{\rm Iterate1} - \alpha_{\rm ELBM}||$.}
\label{alphaIter1tab}
\end{table}

\section{Exact solution to the entropic lattice ES--BGK model} \label{esbgk}

The ES--BGK model proposed by \citet{holway1965kinetic} overcomes the restriction on the Prandtl number ($\rm Pr$) in BGK collision models without compromising the conceptual simplicity. This model employs a quasi-equilibrium state $f^{\rm QE}$ instead Maxwellian in the collision term. The quasi-equilibrium state is an anisotropic Gaussian distribution that reduces to a Maxwellian at the equilibrium.
The continuous $H$ theorem for this model was proved by \citet{andries2000gaussian}. In this section, we extend the discrete $H$ theorem to the lattice ES--BGK model and derive the exact solution for the path length.

\subsection{Lattice ES--BGK model}
The collision term for the lattice ES--BGK collision model reads as \citep{meng2013lattice}
\begin{equation}
\bm f^{*}({\bm x},t) = \bm f({\bm x},t) + \alpha\beta [\tilde{\bm f}^{\rm QE} - \bm f({\bm x},t) ],
\label{esbgkalpha}
\end{equation}
where $\beta = \Delta t/(2\tau_1+\Delta t),$ and the viscosity $\nu$ is related to the relaxation time $\tau_1$ by $ \nu = \tau_1 \theta \,{\rm Pr}$ \citep{kolluru2022reduced}. 
In Eq. \eqref{esbgkalpha}, $\alpha$ is the path length which is equal to 2 in the standard case, and is found by solving Eq. \eqref{entropy_decrease} for the entropic lattice ES--BGK model.
The discrete quasi-equilibrium distribution $\tilde{f}^{\rm QE}$ is found as the minimizer of the the discrete $H$ function under the constraints of mass and momentum being conserved with the pressure tensor given by
\begin{align}
P_{\alpha\beta} \left( \tilde{\bm f}^{\rm QE} \right) &= \rho u_\alpha u_\beta + \rho \theta \delta_{\alpha\beta} + \frac{1-1/{\rm Pr}}{1+\Delta t/(2\tau_1{\rm Pr})} \sigma_{\alpha\beta} \left( \bm f \right). 
\end{align}
Solving the minimization problem, one obtains
\begin{align}
\tilde{f}_i^{\rm QE} = \rho \exp \left( -\mu - {\zeta_\kappa c_{i\kappa}} - \gamma_{\alpha\beta}\sigma_{\alpha\beta} \right),
\end{align}
where $\mu, \zeta_\kappa, \gamma_{\alpha\beta}$ are the Lagrange multipliers associated with the mass, momentum, and pressure tensor respectively. The Lagrange multipliers are calculated by performing a perturbation expansion around the equilibrium state as in Ref. \cite{ansumali2007quasi}. 

\subsection{Exact solution for the path-length}
Following the procedure as detailed in Appendix \ref{derivationdeltaH}, the Eq. \eqref{entropy_decrease} for the lattice ES--BGK model is rewritten as
\begin{align}
\begin{split}
\Delta H = \left< f , ( 1 + \hat z) \ln (1+\hat z) \right> - \alpha\beta \left< f, z \ln (1+z) \right> \\+ \frac{\alpha\beta}{ {\rm Pr} } \frac{1+\delta t/(2\tau_1)}{1+\delta t/(2\tau_1{\rm Pr})} \gamma_{\alpha\beta}  \sigma_{\alpha\beta}(f),
\end{split}
\end{align}
where $\hat z = \alpha\beta z, z = \tilde{f}^{\rm QE}/f - 1$. 
The Lagrange multipliers are evaluated numerically, however, using a series expansion it can be shown that the last term in the above equation can be approximated as 
\begin{equation}
\gamma_{\alpha\beta} = -m \frac{ \sigma_{\alpha\beta} } {\rho \theta^2}, \quad m = \frac{1}{2} \,{\rm for} \, \alpha=\beta, \quad m= 1 \,{\rm otherwise}.
\end{equation}
It is seen that the last term is negative definite hence it contributes only to the entropy production. Thus, the analytical expression for the path length remains the same with equivalent features as Sec. \ref{eelbmsec}.

\subsection{Rayleigh-B\'enard convection}

Rayleigh-B\'enard convection is a well-studied model of natural convection and is considered a classical benchmark for thermal models \citep{shan1997simulation}. 
The domain consists viscous fluid confined between two thermally well-conducting parallel plates. 
The plates are kept at a distance $L$ with the bottom plate maintained at higher temperature $\theta_{\rm bottom}$ and the top plate is kept at a lower temperature $\theta_{\rm top}$. 
The flow is induced by the unstable density gradients in the presence of a gravitational field \citep{atif2018}. 
The dynamics of the Rayleigh-B\`enard convection is characterized by two non-dimensional numbers: the Rayleigh number and the Prandtl number.
The Prandtl number is a property of the fluid (${\rm Pr} = \nu/\alpha_T$) whereas the Rayleigh number (${\rm Ra}$) is defined as
\begin{equation}
{\rm Ra} = \frac{ { g}\hat\beta \Delta \theta L^3}{\nu \alpha_T},
\end{equation}
where ${ g}$ is the gravity, $\hat\beta = 
-1/\rho(\partial \rho/ \partial T)_P$ is the thermal expansion 
coefficient, $\Delta \theta = \theta_{\rm bottom} - \theta_{\rm top}$ is the temperature difference between the two 
walls,
$\nu$ is the kinematic viscosity, and $\alpha_T$ is the thermal diffusivity.  

\begin{figure}
\centering
 \includegraphics[width=0.45\textwidth]{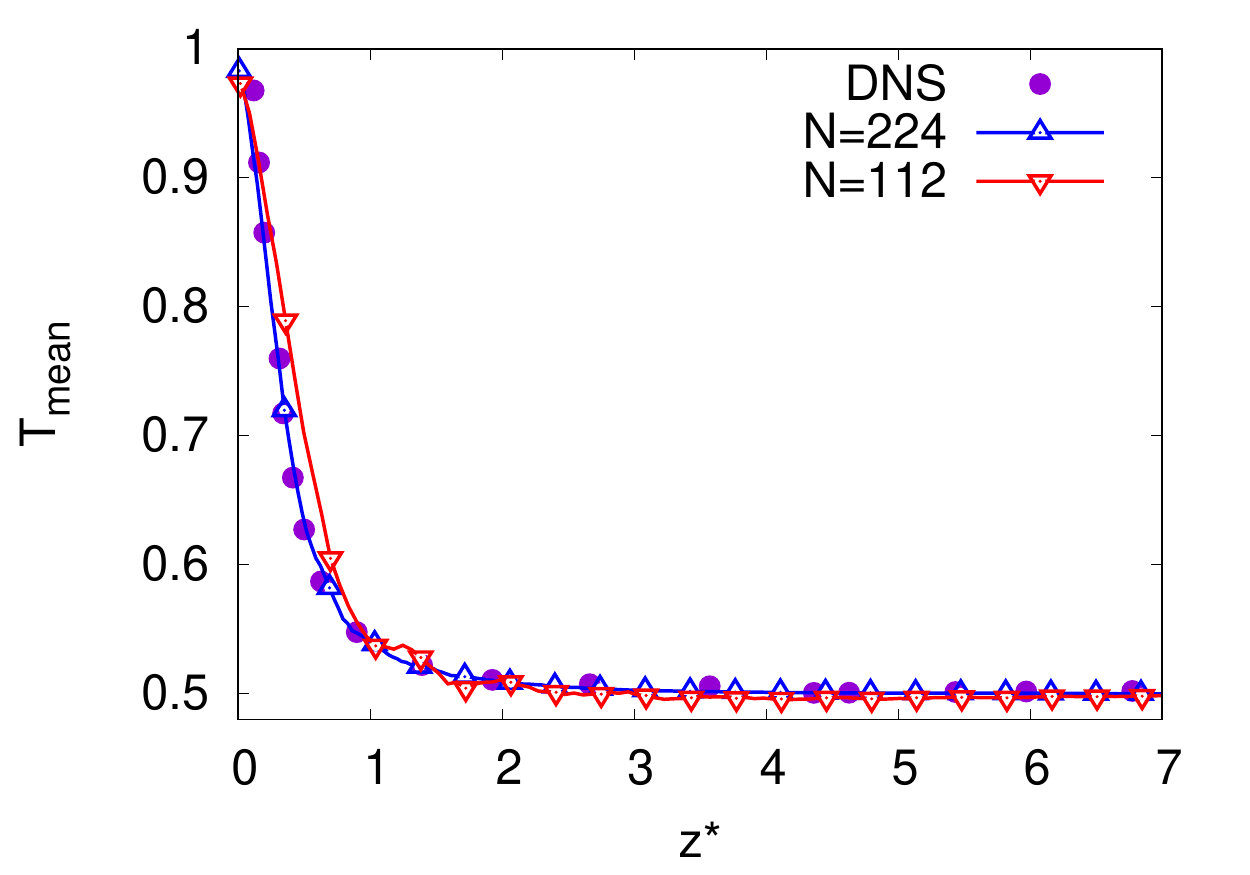}
 \caption{Time-averaged mean horizontal temperatures for Rayleigh-B\`enard convection at ${\rm Ra}=1 \times 10^7$ and ${\rm Pr}=0.71$ compared with the DNS profile from Ref.\cite{togni2015}. Here, the nondimensional vertical coordinate $z^*$ is $z\, {\rm Nu}/L$ }.
 \label{rayberfig}
\end{figure}

In this section, we simulate the turbulent Rayleigh-B\'enard convection at ${\rm Ra} = 1.0 \times 10^7$ and ${\rm Pr}=0.71$ on a grid of size $2N\times 2N\times N$ with $N=112$ and $N=224$.
The exact solution for the path length as derived in the preceding section is used with Eq. \eqref{esbgkalpha} as the collision model.
The numerical simulations are performed using the 67 velocity crystallographic lattice \cite{atif2018} with $\theta_{\rm bottom}=1.02\theta_0$ and $\theta_{\rm top} = 
0.98\theta_0$.
Constant temperature boundary conditions at the top and the bottom walls were 
imposed and periodic boundary conditions were applied in the horizontal directions. 
We  calculate the Nusselt number and time-averaged horizontal mean of nondimensional temperature $T = (\theta - \theta_{\rm top})/\Delta \theta$. The calculated Nusselt number is $13.4$ with $N=112$ and $15.3$ with $N=224$, whereas that reported by the direct numerical simulation (DNS) of Ref. \cite{togni2015} is 15.59. In Fig. \ref{rayberfig} we compare the time-averaged mean horizontal temperature with the DNS data and observe a good match. It can be seen that as expected the temperature rises rapidly close to the wall and obtains a uniform profile in the bulk. Hence, it can be concluded that the exact solution to the path length extends the unconditional numerical stability to non-unity Prandtl number heat transfer simulations too.  

% A detailed quantitative analysis of high Ra convection will be presented elsewhere.
\begin{figure}
 \includegraphics[width=0.45\textwidth]{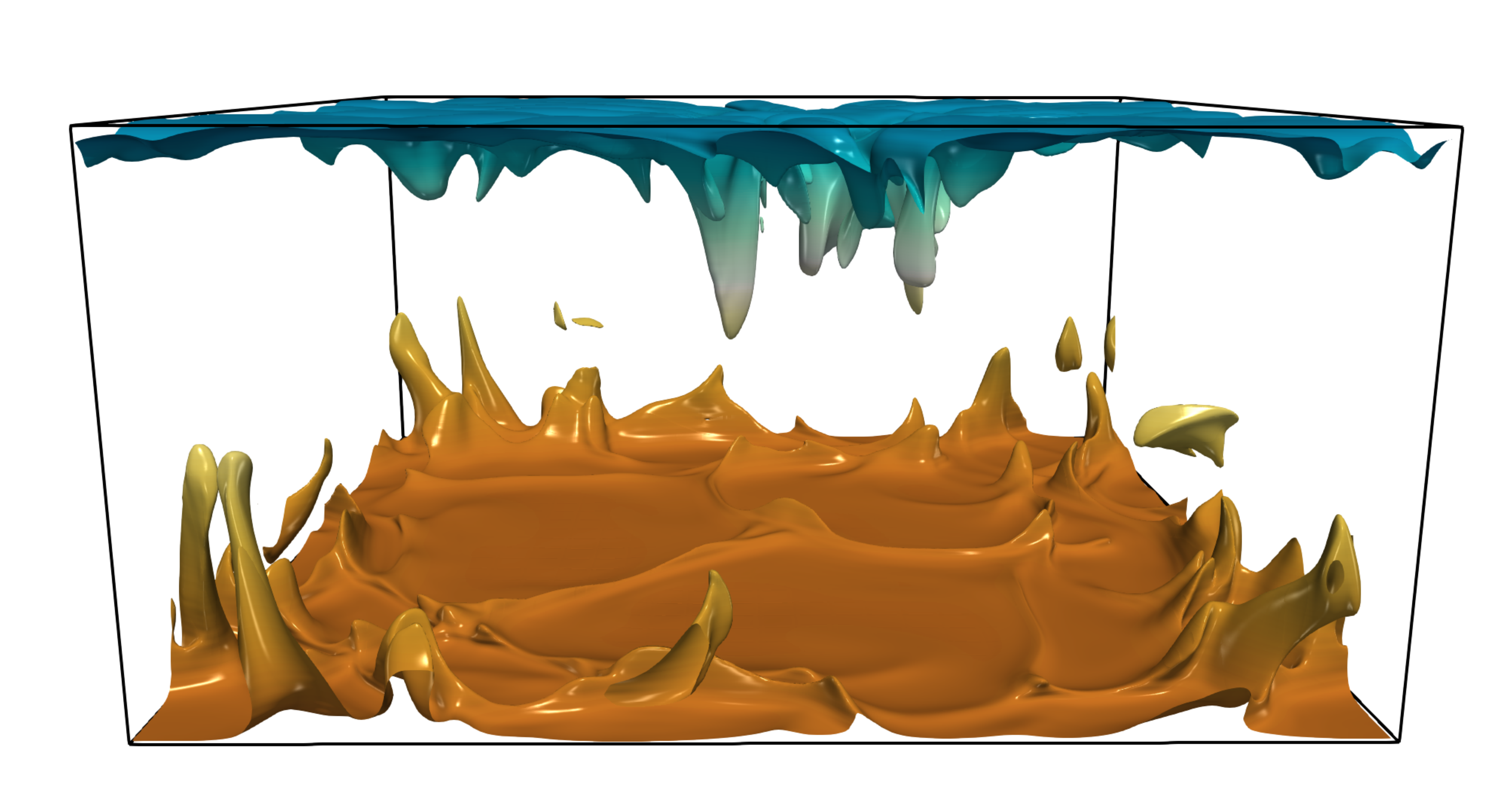}
 \includegraphics[width=0.4\textwidth]{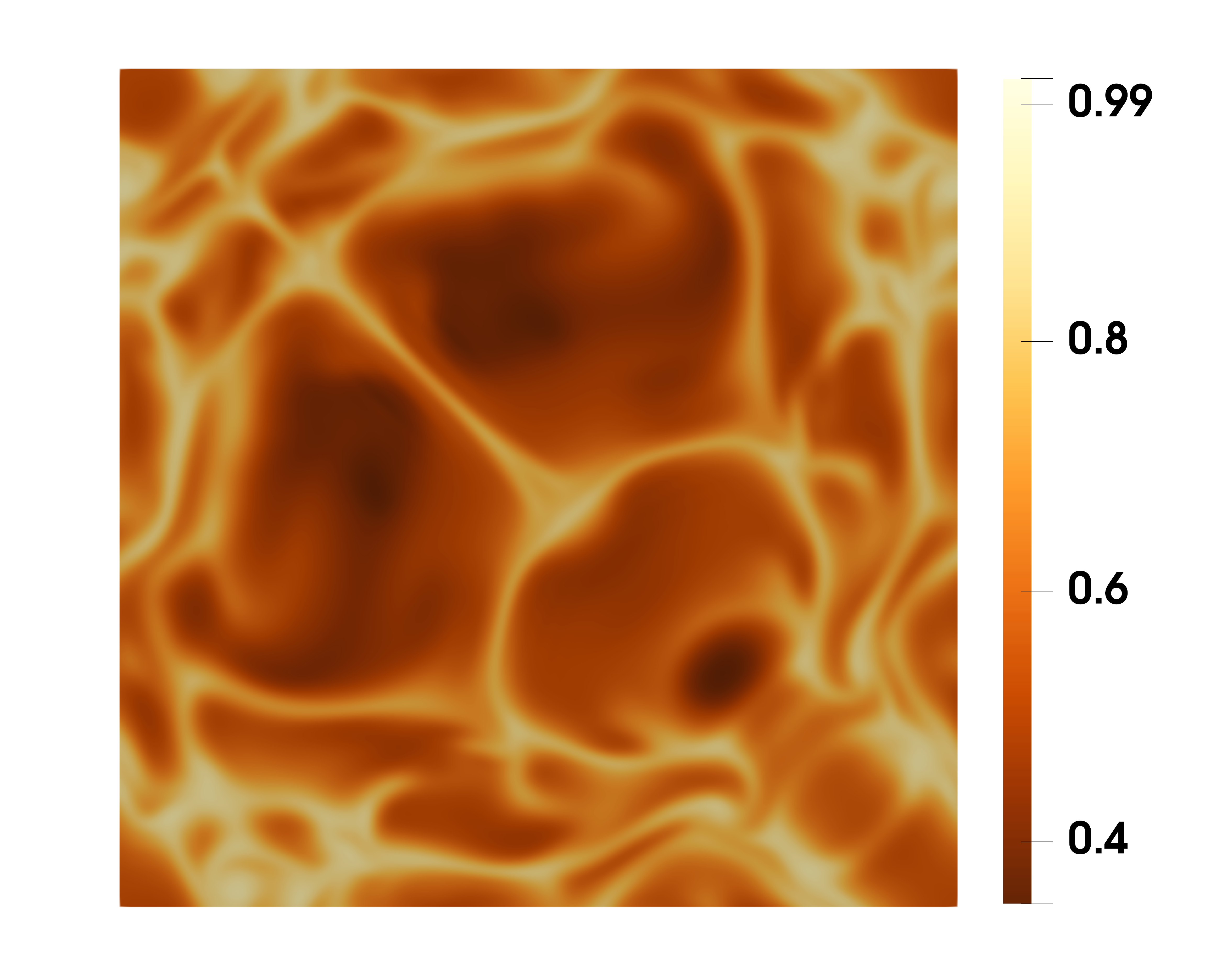}
 \includegraphics[width=0.4\textwidth]{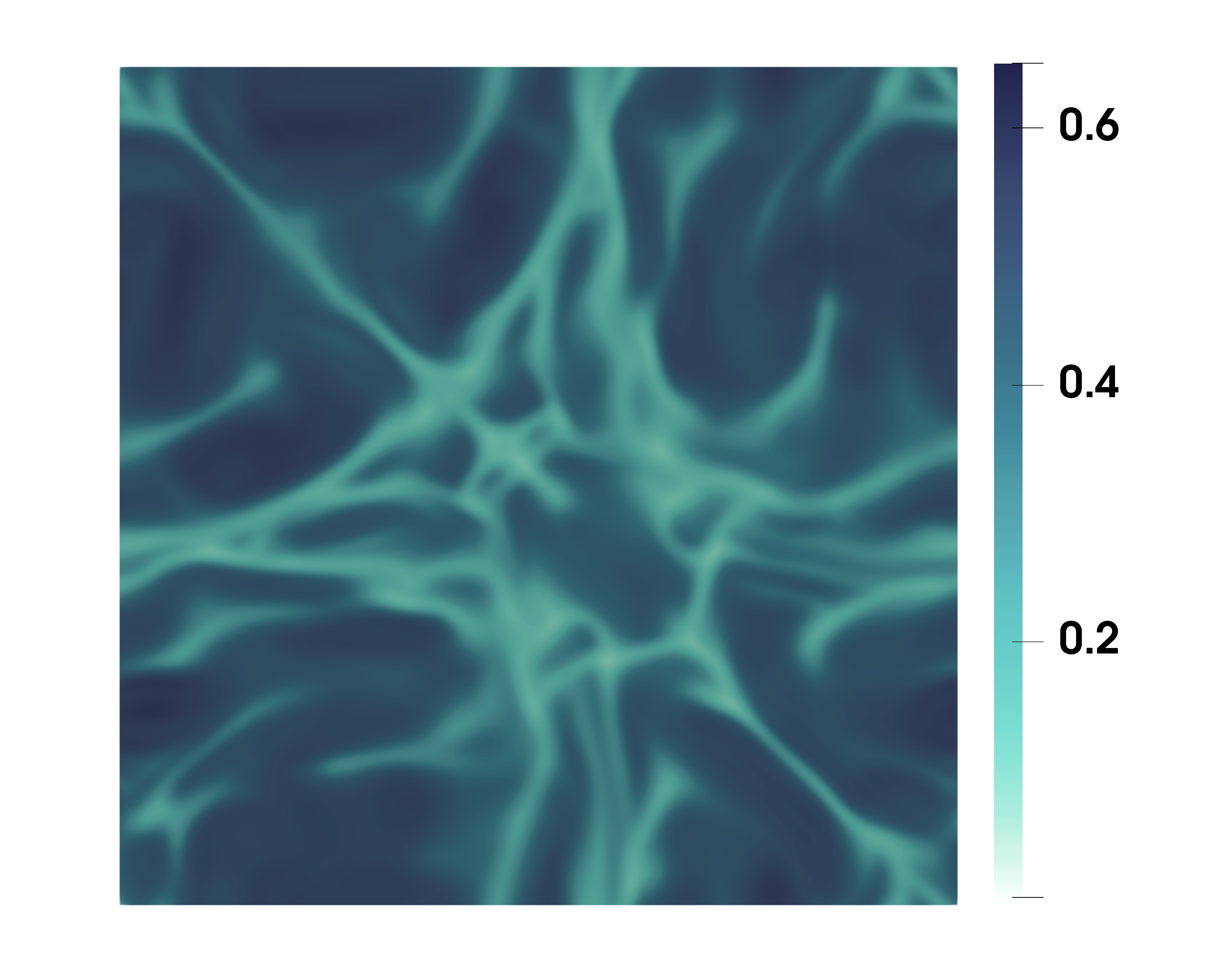}
 \caption{Iso-temperature contours (top) for Rayleigh-B\`enard convection at nondimensional temperatures 0.3, 0.7. The mid and bottom figures visualize the temperature field at horizontal slices close to the two walls. }
 \label{raybertheta}
\end{figure}

\section{Entropic route to modeling the subgrid viscosity} \label{subvisc}

The entropic LBM has been interpreted as an implicit subgrid model of turbulence \cite{karlin2003entropic}. 
The modification to the path length $\alpha$ due to the compliance with the $H$ theorem can be understood as a turbulent viscosity correction at the macroscopic scale.
Several studies have analyzed the form of the viscosity correction and found similarities to the Smagorinsky's model for viscosity correction \cite{deville2008towards,tauzin2021inertial}.
In this section, we derive the subgrid model corresponding to the exact path length.
From the Chapman-Enskog expansion the effective kinematic viscosity $\nu$ due to the entropic collision term is found as
\begin{equation}
\nu = \theta \tau= \theta\Delta t \left( \frac{1}{\alpha \beta} - \frac{1}{2} \right) .
\end{equation}
The viscosity correction $\nu_T$ is defined as $\nu_T= \nu - \nu_0$, where $\nu_0$ is the viscosity corresponding to the BGK path length $\alpha=2$, and is obtained as
\begin{equation}
\nu_T = \frac{\theta\Delta t}{2\alpha\beta} \left( 2-{\alpha} \right).
\label{nuT}
\end{equation}
It is seen from the above expression that the path length $\alpha$ dictates whether the viscosity correction is positive or negative. 
A path length smaller than 2 implies an increment in the viscosity which in turn smoothens the gradients, whereas, a path length larger than 2 corresponds
to reduction in the viscosity which sharpens the gradients \citep{karlin2015entropy}.
Thus, the entropic LBM permits backscatter of energy from the subgrid scales to the resolved scales too.

We now evaluate the viscosity correction in terms of the macroscopic moments.
For this purpose we consider the path length from Eq. \eqref{alphahigherlimit} in assuming small departure from equilibrium, i.e., $x_i \rightarrow 0, \, \Delta t \gg \tau$, and interpret $\left< \cdot \right>$ as a continuous integral.
Substituting Eq. \eqref{alphahigherlimit} in Eq. \eqref{nuT} the turbulent viscosity correction $\nu_T$ is found as
 \begin{equation}
\nu_T= \frac{\theta\Delta t}{2} \frac{\Theta}{6+\Theta}, \, {\rm where} \, \quad \Theta = \frac{\left<f,x^3 \right>}{\left<f,x^2 \right>}.
\end{equation} 
From Grad's $13$ moment representation one can write the approximation $f = f^{\rm MB} \left( 1 + \Omega \right)$,
where 
 \begin{equation}
\Omega = \frac{\sigma_{ij}\xi_i\xi_j}{2 p \theta} -\frac{q_k \xi_k}{p\theta}\left(
1 - \frac{\xi^2}{5\theta}\right),
\end{equation} 
$f^{\rm MB}$ is the Maxwell-Boltzmann distribution, $\xi_i$ is the peculiar velocity, $p$ is the pressure, $\theta$ is the temperature, $\sigma_{ij}$ is the traceless part of symmetric stress tensor and $q_k$ is the heat flux. 
% We further approximate the two  as
% \begin{align}
% \begin{split}
% \left<f,x^2 \right> &= 
% \left< f^{\rm MB}, \frac{\Omega^2}{ 1 + \Omega } \right> \approx \left< f^{\rm MB}, {\Omega^2}{ \left( 1 - \Omega \right) } \right> \\
% \left<f,x^3 \right> &= 
% \left< f^{\rm MB}, -\frac{\Omega^3}{ \left(1 + \Omega\right)^2 } \right> \approx \left< f^{\rm MB}, -{\Omega^3}{ \left( 1 - 2\Omega \right) } \right>\\
% \end{split}
% \end{align}
Thereafter, the leading terms of the two terms appearing in $\Theta$ are evaluated as
\begin{align}
\begin{split}
\left<f,x^2 \right> &= \frac{1}{2p\theta}\sigma_{kl}\sigma_{lk} + {\cal O}(\sigma^3),\\
\left<f,x^3 \right> &= -\frac{3}{p^2\theta}\sigma_{kl}\sigma_{lm}\sigma_{mk} +{\cal O}(\sigma^4),
\end{split}
\end{align}
 where assuming a small change in temperature ${\cal O}(q^2)$ terms have been ignored.  
 Substituting $\sigma_{ij} = \rho \tau \theta S_{ij}$, $\bm S$ being the strain rate tensor we find the viscosity correction as  
\begin{equation}
\nu_T = -{\tau\theta} \, \frac{\Delta t}{2}  \frac{S_{ij}S_{jk}S_{ki}}{  S_{mn} S_{nm} - \tau S_{ab} S_{bc} S_{ca} }.
\end{equation}
It should be noted that for very fine grid resolutions ($\Delta t \rightarrow 0$) the viscosity correction vanishes.
Similar expressions for the turbulent viscosity have also been derived in Refs. \cite{deville2008towards,tauzin2021inertial}.
The above expression for turbulent viscosity is similar to Smagorinsky's model where the turbulent viscosity $\nu_T$ is
\begin{equation}
\nu_T = (C_S \Delta)^2 \sqrt{S_{ij}S_{ji}},
\end{equation}
where $C_S$ is Smagorinsky's constant, in that, both scale like the strain rate tensor and is also distinct from it because of emergence of the third invariant of the symmetrized strain rate tensor \citep{smagorinsky1965numerical,deardorff1970numerical}.

\section{Conclusion}

% To conclude, this new exact solution is a step forward in the theoretical development of ELBM. 
In this paper, we present in detail the methodology to construct exact solutions to the path length in the entropic lattice Boltzmann method.
This methodology can be extended to derive more accurate expressions, however, we find that $\alpha_{\rm Higher}$ is sufficient for hydrodynamic applications. 
The more dissipative solution $\alpha_{\rm Lower}$ could also be employed to model viscous flows in the vicinity of walls and can also be used as a good guess for the iterative solution.
We have demonstrated that $\alpha_{\rm Higher}$ shows no appreciable difference from the iterative solution by studying the macroscopic behaviour of a few canonical setups.
We have also extended the exact solution to lattice ES-BGK model for nonlinear numerical stability in non-unitary Prandtl heat transfer scenarios. 

%In Appendix \ref{hTheoremextra}, the discrete time $H$ theorem is proved for the quasi-equilibrium model, which extends the second law to discrete dynamics for non-unitary Prandtl number flows.
%  Furthermore, this essentially entropic LBM provides an important first step in providing statistical mechanics route to subgrid scale modeling.  
%  For example, using discrete entropic space-time dynamics for Boltzmann BGK equation, we have shown that 
%   the  correction to viscosity $\nu_T$  is 
%\begin{equation}
%\nu_T = -{\tau\theta} \, \frac{\Delta t}{2}  \frac{S_{ij}S_{jk}S_{ki}}{  S_{mn} S_{nm} },
%\end{equation}
% where $S_{ij}$ is the strain rate tensor \citep{deville2008towards,chikatamarla2006entropic}. This  emergence of the third invariant of symmetrized strain rate tensor is distinct from
%   Smagorinsky's model for turbulent viscosity. 
%Though physically appealing \citep{meneveau2011lagrangian}, further detailed numerical and theoretical analysis of current framework  are needed to establish usefulness of this approach for theoretical subgrid modeling. 
%  Finally, we highlight the fact that entropic formulation of continuous (in velocity space)  BGK model provides 
%  a new discrete dynamical system analogue of Boltzmann dynamics.  Thus, a Boltzmann like framework extends the second law to discrete dynamical systems too.

\appendix

\section{Derivation of $\Delta H$} \label{derivationdeltaH}

In this section, we derive the expression for $\Delta H = H[\bm f^{\rm mirror}]-H[\bm f]$. 
We begin by using the form of $H$ [Eq. \eqref{hdef}] to obtain
\begin{align}
\begin{split}
&H[\bm f^{\rm mirror}]-H[\bm f] = \left<f^{\rm mirror},\log\frac{f^{\rm mirror}}{w} \right> - \left<f,\log\frac{f}{w} \right>. 
\end{split}
\end{align}
Substituting $\bm f^{\rm mirror}$ from Eq. \eqref{fmirrordef} in the above equation yields
\begin{align}
\begin{split}
&H[\bm f^{\rm mirror}]-H[\bm f] \\&= \left<f+\alpha( f^{\rm eq} - f ),\log\frac{f+\alpha( f^{\rm eq} - f )}{w} \right> - \left<f,\log\frac{f}{w} \right>.
\end{split}
\end{align}
Substituting $x$ from Eq. \eqref{xdef} in the above equation one obtains
 \begin{align}
 \begin{split}
&H[\bm f^{\rm mirror}]-H[\bm f] 
\\&=\left<f(1+\alpha x),\log\frac{ f(1+\alpha x)}{w} \right> - \left<f,\log\frac{f}{w} \right>
% \\&=\left<f(1+\alpha x),\log { (1+\alpha x)} \right> + \left<f,(1+\alpha x)\log \frac{ f}{w} \right> - \left<f,\log\frac{f}{w} \right>
\\&=\left<
 f\left( 1+  \alpha x \right), 
\log{  \left( 1+  \alpha  x \right)} \right> 
-\alpha   \left< fx, \log \frac{w}{f} \right>.
\label{deltaH}
\end{split}
\end{align}
Now substituting $w_i$ from Eq. \eqref{entropiceqdef} one obtains
 \begin{align}
 \begin{split}
&H[\bm f^{\rm mirror}]-H[\bm f] \\&=\left< f\left( 1+  \alpha x \right), \log{  \left( 1+  \alpha  x \right)} \right> \\&-\alpha   \left< fx, \log \frac{f^{\rm eq}  \exp({\mu + \zeta_\kappa c_{i\kappa} + \gamma c_i^2})}{f} \right>\\
&= \left<  f\left( 1+  \alpha  x \right),  \log{  \left( 1+  \alpha  x\right)} \right> -\alpha  \left< fx, \log(1+x) \right>
\\&- \underline{\alpha \lambda \left<f, x   \right>}
- \underline{ \alpha \zeta_\kappa  \left<f, x  c_{\kappa} \right>}
- \underline{ \alpha \gamma \left<f, x c^2\right>},
\end{split}
\end{align}
where we have substituted $f_i^{\rm eq}/f_i = 1+x_i$ and the underlined terms are zero due to moments invariance, i.e.,
\begin{align}
\begin{split}
\left<f,x \right> = \sum_i ( f_i^{\rm eq} - f_i ) = \rho -\rho=0,\\
 \left<f,x c_{\kappa} \right>  = \sum_i ( f_i^{\rm eq}  c_{i\kappa} - f_i  c_{i\kappa} ) = \rho u_\kappa-\rho u_\kappa=0,\\
\left<f,x c^2 \right> = \sum_i ( f_i^{\rm eq}  c^2_{i} - f_i  c^2_{i} ) = \rho e -\rho e =0.
\end{split}
\end{align}
Thus, we obtain
 \begin{align}
 \begin{split}
&H[\bm f^{\rm mirror}]-H[\bm f] \\&
= \left<  f\left( 1+  \alpha  x \right),  \log{  \left( 1+  \alpha  x\right)} \right> -\alpha  \left< fx, \log(1+x) \right> \\&
= \left<  f, \left( 1+  \alpha  x \right) \log{  \left( 1+  \alpha  x\right)} \right> -\alpha  \left< f, x\log(1+x) \right>.
\label{deltaH2}
\end{split}
\end{align}

\section{Bounds on the logarithm} \label{logbounds}

% Before deriving the analytical expression for the path length, we list some nonnegative functions obtained 
% by exploiting the lower order bounds on the logarithm \citep{atif2017}:
%  \begin{align}
% G_1(y) = -(1+y)\log(1+y) + y + \frac{y^2}{2} - \frac{y^3}{2} > 0 \qquad &y\in(-1,0), \label{G1} \\
% G_2(y) = -(1+y)\log(1+y) +  y + \frac{y^2}{2}  \geq 0 \qquad &y\in [0, \infty), \label{G2}\\
%  G_3(y) = y \log(1+y) - \frac{2y^2}{2+y} \geq 0 \qquad &y\in (-1, \infty). \label{G3}
%  \end{align}
In this section, we list a few positive definite functions along with their domain of validity.
In the interval $y\in(-1,0)$, using using the Taylor series expansion of the logarithm we define
\begin{align}
G_1(y) &= (1+y)\left[-\log(1+y) + y - \frac{y^2}{2} \right] > 0,
\label{G1bound}
\end{align}
\begin{align}
G_4(y) &= (1+y) \Bigg[ -\log(1+y) + y - \frac{y^2}{2} + \frac{y^3}{3} - \frac{y^4}{4} \nonumber\\&+ \frac{y^5}{5} \Bigg]> 0 \label{G4bound}.
 \end{align}

Next, we exploit the integral definition of $\log(1+y)$, i.e.,
\begin{equation}
\log(1+y) = \int_0^y  \frac{1}{1+z} dz,
\label{logDef}
\end{equation}
and evaluate it using Gauss-Legendre and Newton-Cotes quadrature rules.
As the integrand is an $2n$-convex function, i.e., its even ($2n$) derivatives are positive, the error due to the approximations are sign-definite, hence these approximations can be used to construct upper and lower bounds on $\log(1+y)$.

Evaluating the integral in Eq. \eqref{logDef} via Gauss-Legendre quadratures, one obtains 
\begin{align*}
\mathcal{I}_{\rm GL}^{(1)}(y)&=\frac{2y}{(2+y)} ,  \\
\mathcal{I}_{\rm GL}^{(2)}(y)&=\frac{6y + 3y^2}{6+6y+y^2}, \\
\mathcal{I}_{\rm GL}^{(3)}(y)&=\frac{60y+60y^2+11y^3}{60+90y+36y^2+3y^3},
\end{align*}
where $\mathcal{I}_{\rm GL}^{(n}$ is the intergral evaluated using $n^{\rm th}$-order Gauss-Legendre quadrature. 
Similarly, evaluating the integral in Eq. \eqref{logDef} via Newton-Cotes quadratures, one obtains \cite{khattri}
\begin{align*}
\mathcal{I}_{\rm NC}^{(1)}(y)&=\frac{y}{2} \left[1+ \frac{1}{1+y}\right], \\
\mathcal{I}_{\rm NC}^{(2)}(y)&=\frac{y}{6} \left[1 + \frac{8}{2+y} + \frac{1}{1+y} \right],\\
\mathcal{I}_{\rm NC}^{(4)}(y)&=\frac{y}{90} \left[7 + \frac{128}{4+y} + \frac{48}{4+2y} + \frac{128}{4+3y} + \frac{7}{1+y} \right],
\end{align*}
where $\mathcal{I}_{\rm GL}^{(n}$ is the intergral evaluated using $n^{\rm th}$-order Newton-Cotes quadrature.

In the interval $y\in [0, \infty)$, exploiting the sign-definiteness of the errors we define
\begin{align}
G_2(y) = (1+y)\left[-\log(1+y) + \mathcal{I}_{\rm NC}^{(1)} \right] \geq 0, 
\label{G2bound}
\end{align}
\begin{align}
G_5(y) = (1+y)\left[-\log(1+y) + \mathcal{I}_{\rm NC}^{(3)} \right] \geq 0,
\label{G5bound}
\end{align}
and in the interval $y\in (-1, \infty)$ we define 
\begin{align}
G_3(y) = y \left[ \log(1+y) - \mathcal{I}_{\rm GL}^{(1)} \right] \geq 0,
\label{G3bound}
\end{align}
\begin{align}
G_6(y) = y \left [\log(1+y) - \mathcal{I}_{\rm GL}^{(3)} \right] \geq 0.
\label{G6bound}
\end{align}
 
The functions $G_1(y), G_2(y), G_3(y)$ form loose bounds bounds on the logarithm, whereas, $G_4(y), G_5(y), G_6(y)$ provide sharp bounds on it.

\section{Derivation of the higher-order solution} \label{higherordersoln}

Following the same methodology  as Section \ref{lowerordersoln}, we add and subtract the same terms from Eq. \eqref{deltaH} to obtain
\begin{align}
\begin{split}
\Delta H &= H^{(B)}+ \alpha\beta\hat H(\alpha),
\end{split}
\label{deltaHtight}
\end{align}
where
\begin{align}
H^{(B)} = -\Big<   f, {G_6(\alpha\beta x) }_{} \Big>_{\Omega^{-}} 
- \Big<  f, { G_7(\alpha\beta x) }_{} \Big>_{\Omega^{+}}
 { -\alpha\beta G_8(x) }_{} \leq 0,
\end{align}
is nonpositive and contributes to the entropy production, and 
\begin{align}
\begin{split}
\hat H(\alpha) =-\left<   f, \frac{\alpha^2 \beta^2x^3}{6} - \frac{\alpha^3 \beta^3x^4}{12} + \frac{\alpha^4 \beta^4 x^5}{20} - \frac{\alpha^5 \beta^5 x^6}{5} \right>_{\Omega^{-}} \\
+ \left<f, \frac{\alpha \beta x^2}{2}  \right>
- \bigg<f, \frac{2\alpha^2 \beta^2 x^3}{15}\bigg(
\frac{2}{4+\alpha \beta x} 
 + \frac{1}{4+2\alpha \beta x} 
\\+ \frac{2}{4+3\alpha \beta x}   \bigg) \bigg>_{\Omega^{+}}   
-  \left< f, \frac{60x^2+60x^3+11x^4}{60+90x+36x^2+3x^3} \right>.
\end{split}
\label{sexticPolynomial}
\end{align}
The above equation has at least one positive root as $\hat H(0) < 0 < \hat H(\infty)$ which can be found using any numerical method. In order to preserve the computational efficiency of the method we solve the above equation by converting it into a quadratic in $\alpha$.

\subsection{Solving the higher degree polynomial} \label{quadtoquint}

\begin{figure}
   \centering
    {\includegraphics[width=0.35\textwidth]{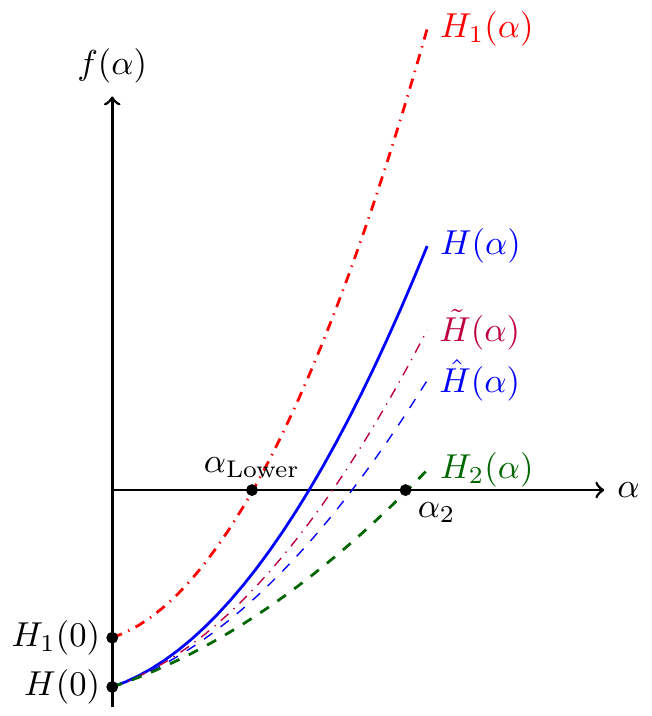}}
    \caption{Behaviour of Eqs. \eqref{dissipativescheme},\eqref{sexticPolynomial},\eqref{quinticPolynomial},\eqref{qaud_tight},\eqref{gaplhah0} near the positive root.}
\label{roots}
\end{figure}

In this section, we solve Eq. \eqref{sexticPolynomial} by converting it to a quadratic.
This conversion to quadratic is performed by extracting negative terms from the Eq. \eqref{sexticPolynomial}. 
The extracted terms then contribute to the entropy production $ H^{(B)}$.
As stated earlier, the Eq. \eqref{sexticPolynomial} has a positive root since $\hat H(0) < 0 < \hat H(\infty)$.
We assume that upper and lower bounds on the root $\alpha$ exist. 
A suitable choice for the lower bound is $\alpha_{\rm Lower}$,
while the upper bound $h$ will be later evaluated.
Therefore,
$\alpha_{\rm Lower} < \alpha < h$. Converting $\hat H(\alpha)$ to a quadratic is a two step procedure and is explained in the following subsections.

\begin{widetext}
\subsubsection{Exploiting the lower bound}
Using the lower bound $\alpha_{\rm Lower}$, in Eq. \eqref{sexticPolynomial} we split the term
\begin{align}
\begin{split}
- &\bigg<f, \frac{2\alpha^2 \beta^2 x^3}{15}\bigg( \frac{2}{4+\alpha \beta x}  + \frac{1}{4+2\alpha \beta x} 
+ \frac{2}{4+3\alpha \beta x}   \bigg) \bigg>_{\Omega^{+}}  
\\& 
\equiv - \bigg<f, \frac{2\alpha \beta^2 x^3}{15}\bigg( \frac{2}{\frac{4}{\alpha_{\rm Lower}}+ \beta x}   + \frac{1}{\frac{4}{\alpha_{\rm Lower}}+2 \beta x}+ \frac{2}{\frac{4}{\alpha_{\rm Lower}}+3 \beta x}   \bigg) \bigg>_{\Omega^{+}} 
\\& 
- \bigg<f, \frac{2\alpha \beta^2 x^3}{15}\bigg( 
\bigg\{ \frac{2}{\frac{4}{\alpha}+ \beta x}  - \frac{2}{\frac{4}{\alpha_{\rm Lower}}+ \beta x}  \bigg\} +\bigg\{ \frac{1}{\frac{4}{\alpha}+2 \beta x} - \frac{1}{\frac{4}{\alpha_{\rm Lower}}+2 \beta x}  \bigg\} + \bigg\{ \frac{2}{\frac{4}{\alpha}+3 \beta x} - \frac{2}{\frac{4}{\alpha_{\rm Lower}}+3 \beta x}   \bigg\} \bigg) \bigg>_{\Omega^{+}} ,
\label{ksplit}
\end{split}
\end{align}
where each term in curly braces is positive (as $\alpha_{\rm Lower} < \alpha$) thereby making the second term negative.
Here, recognizing that the negative term contributes to the entropy production $ H^{(B)}$, we obtain the quintic polynomial $\tilde H(\alpha)$,
\begin{align}
\begin{split}
\tilde H(\alpha) &= -\alpha^2 \beta^2 \left<   f, \frac{ x^3}{6} - \frac{\alpha \beta x^4}{12} + \frac{ \alpha^2 \beta^2 x^5}{20} - \frac{ \alpha^3 \beta^3 x^6}{5} \right>_{\Omega^{-}}
+ \alpha \bigg[ \bigg<f, \frac{ x^2}{2}  \bigg>
- \bigg<f, \frac{ 2 \alpha_{\rm Lower} \beta^2 x^3}{15}\bigg(
\frac{2}{ 4+\alpha_{\rm Lower} x } \\& + \frac{1}{ 4+2 \alpha_{\rm Lower} x } 
+ \frac{2}{ 4 +3 \alpha_{\rm Lower} x}   \bigg) \bigg>_{\Omega^{+}} \bigg]
-\left< f, \frac{60x^2+60x^3+11x^4}{60+90x+36x^2+3x^3} \right>.
\end{split}
\label{quinticPolynomial}
\end{align}
Essentially, while converting $\hat H(\alpha)$ to  $\tilde H(\alpha)$, we have shifted the negative definite terms in Eq. \eqref{ksplit} to the entropy production, 
hence, the curve for $\tilde H(\alpha)$ lies above $\hat H(\alpha)$ (see Fig. \ref{roots}). 
It follows that an upper bound on the root of $\hat H(\alpha)$
will also serve as the upper bound for the root of $\tilde H(\alpha)$. 

\subsubsection{Exploiting the upper bound}

Using the upper bound $h$, in Eq. \eqref{quinticPolynomial} we split the term
\begin{align}
&-\left<   f, \frac{\alpha^2 \beta^2x^3}{6} - \frac{\alpha^3 \beta^3x^4}{12} + \frac{\alpha^4 \beta^4 x^5}{20} - \frac{\alpha^5 \beta^5 x^6}{5} \right>_{\Omega^{-}} \equiv -\alpha^2 \beta^2 \left<   f, \frac{ x^3}{6} - \frac{ h \beta x^4}{12} + \frac{ h^2 \beta^2 x^5}{20} - \frac{ h^3 \beta^3 x^6}{5} \right>_{\Omega^{-}} \nonumber\\&
 -\alpha^2 \beta^2 \left<   f, - \frac{ (\alpha-h) \beta x^4}{12} + \frac{ (\alpha^2-h^2) \beta^2 x^5}{20} - \frac{(\alpha^3- h^3) \beta^3 x^6}{5} \right>_{\Omega^{-}},
\label{hsplit}
\end{align}
where the second term is negative, due to $x_i<0, x_i \in \Omega^- \, {\rm and}\, \alpha<h$. 
Now, substituting Eq. \eqref{hsplit} into Eq. \eqref{quinticPolynomial} and again recognizing that the negative terms contribute to the entropy production $ H^{(B)}$, we obtain the quadratic $H(\alpha)$.

\end{widetext}

It remains to specify the upper bound $h$.
For this we consider the quadratic equation  $H_2(\alpha) = H(\alpha)|_{h=0}$, 
\begin{align}
\begin{split}
H_2(\alpha) &= -\alpha^2 a_2 + \alpha b - c,
\end{split}
\end{align}
\begin{align}
\begin{split}
a_2 = \beta^2 \left<   f, \frac{ x^3}{6} \right>_{\Omega^{-}},
\end{split}
\label{a2def}
\end{align}
whose positive root is $\alpha_2$. Therefore, $H_2(\alpha_2) = 0$ and
 \begin{align}
H({\alpha_2}) = \alpha_2^2 \beta^2 \left<   f,  \frac{ h \beta x^4}{12} - \frac{ h^2 \beta^2 x^5}{20} + \frac{ h^3 \beta^3 x^6}{5} \right>_{\Omega^{-}} \nonumber\\+ H_2(\alpha_2) > 0.
\end{align} 
As $H(0)<0<H(\alpha_2)$, a root of $H(\alpha)$ lies in the interval $(0,\alpha_2)$ (see Figure \ref{roots}). Hence,
a suitable choice for the upper bound is $h=\alpha_2$.

\section{Implementing the analytical solution} \label{implementing}

The post-collisional populations are found via the routine
\begin{equation}
f_i^* = f_i  +\alpha\beta [ f_i^{\rm eq} - f_i ],
\end{equation}
where the path length $\alpha$ needs to be evaluated at each grid point.
We begin by calculating
\begin{equation}
x_i = \frac{f_i^{\rm eq}}{f_i}-1,
\end{equation}
where $i=1 \rightarrow N$ for a lattice with $N$ discrete velocities. To evaluate a summation on one of the sub-divisions $\Omega^{-}$ or $\Omega^{+}$ we sum over the populations in the concerned subdivision.
For instance, to calculate 
$$a_1 = \left<f, \frac{x^3}{2} \right>_{\Omega^{-}}, \quad b_1 = \left<f, \frac{x^2}{2} \right>,$$ the pseudo-code is:

\begin{algorithm}[H]
\begin{algorithmic}[1]
\State $a_1 = 0, b_1=0$
\For {each integer $i$ in $1$ to $N$}
  \If {$x_i < 0$}
      \State $a_1 = a_1 + f_i * x_i^3/2$
  \EndIf
\State $b_1 = b_1 + f_i * x_i^2/2$
\EndFor
\State Return $a_1$
\end{algorithmic}
\end{algorithm}

To find the path length $\alpha$ we execute the following steps:

\begin{algorithm}[H]
\begin{algorithmic}[1]
    \State Find $|x_i|^{\rm max}$, the $x_i$ with maximum magnitude.
    \If {$|x_i|^{\rm max} < 10^{-3}$}
        \State $\alpha=2$
    \Else
        \State Calculate $a_1, b_1, c_1$ from Eq. \eqref{alphaLowerCoeffs}
        \State Calculate $\alpha_{\rm Lower}$ from Eq. \eqref{alphaLowerSoln} 
        \State Calculate $a_2$ from Eq. \eqref{a2def} and $b,c$ from Eq. \eqref{qaud_tight_coeff}
        \State Calculate $h$, the positive root of the Eq. \eqref{gaplhah0}
        \State Calculate $a,b,c$ from Eq. \eqref{qaud_tight_coeff}
        \State Find $\alpha_{\rm Higher}$, the positive root of Eq. \eqref{qaud_tight}
        \State $\alpha = \alpha_{\rm Higher}$
    \EndIf
\end{algorithmic}
\end{algorithm}

Although, the exact solution to the path length is always found, we need to ensure that the post collisional populations remain positive due to the boundary conditions or in the case of of extremely under-resolved situations.
To this effect, an extra step might be required. 
We again stress that these situations are extremely rare.
The maximum permitted value of the path length such that all the post collisional populations remain positive is $\alpha^{\rm max}$. Therefore,

\begin{algorithm}[H]
\begin{algorithmic}[1]
        \State Find $x_i^{\rm min}$, the smallest $x_i$
        \State Calculate $\alpha^{\rm max} = -1/(\beta x_i^{\rm min})$
        \If {$ \alpha > \alpha^{\rm max}$ }
          \State $\alpha = (1+\alpha^{\rm max})/2$
        \EndIf
\end{algorithmic}
\end{algorithm}

\bibliography{./rd3q67}

\end{document}